\documentclass[journal, onecolumn, twoside]{IEEEtran}

\newif\iffull
\fulltrue

\usepackage{algorithmic, algorithm}
\usepackage[english]{babel}
\usepackage{amssymb}
\usepackage{amsmath}
\usepackage{xcolor}
\usepackage{cite}
\usepackage{graphicx}
\usepackage{amsthm}
\usepackage[hidelinks]{hyperref}
\usepackage{comment}
\usepackage{enumitem}
\setlist[itemize]{leftmargin=*}
\usepackage{orcidlink}
\usepackage{cancel}

\theoremstyle{definition}
\newtheorem{theorem}{Theorem}
\newtheorem{lemma}{Lemma}

\newtheorem{definition}{Definition}
\newtheorem{corollary}{Corollary}
\newtheorem{problem}{Problem}

\newtheorem{example}{Example}

\newcommand{\bfc}{{\boldsymbol c}}

\newcommand{\bfp}{{\boldsymbol p}}
\newcommand{\bfq}{{\boldsymbol q}}

\newcommand{\bfs}{{\boldsymbol s}}

\newcommand{\bfu}{{\boldsymbol u}}
\newcommand{\bfv}{{\boldsymbol v}}
\newcommand{\bfw}{{\boldsymbol w}}
\newcommand{\bfx}{{\boldsymbol x}}
\newcommand{\bfy}{{\boldsymbol y}}
\newcommand{\bfz}{{\boldsymbol z}}

\newcommand{\bfalpha}{{\boldsymbol \alpha}}
\newcommand{\bfbeta}{{\boldsymbol \beta}}

\newcommand{\cC}{\mathcal{C}}

\newcommand{\cK}{\mathcal{K}}

\newcommand{\cO}{\mathcal{O}}
\newcommand{\cP}{\mathcal{P}}

\newcommand{\cS}{\mathcal{S}}

\renewcommand{\Bbb}{\mathbb}

\newcommand{\N}{{\Bbb N}}

\usepackage{tkz-graph}
\usetikzlibrary{arrows}
\usetikzlibrary{positioning}
\usetikzlibrary{calc}
\usepackage{thm-restate}
\usepackage{cleveref}
\usepackage{subcaption}

\title{On the Capacity of DNA Labeling}

\author{
    Dganit~Hanania, Daniella~Bar-Lev, Yevgeni~Nogin, Yoav~Shechtman and Eitan~Yaakobi
    
\thanks{%
Parts of this work were presented at the IEEE International Symposiumon Information Theory (ISIT), Taipei, Taiwan, 2023~\cite{ISITpaper}. 

D. Hanania, D. Bar-Lev and E. Yaakobi are with the Henry and Marilyn Taub Faculty of Computer Science, Technion- Israel Institute of Technology, Haifa 3200003, Israel (e-mail: {dganit, daniellalev, yaakobi}@cs.technion.ac.il). Y. Nogin is with Russel Berrie Nanotechnology Institute, Technion, Haifa 320003, Israel (e-mail: ynogin@campus.technion.ac.il). Y. Shechtman is with Russel Berrie Nanotechnology Institute, with Department of Biomedical Engineering and with Lorry I. Lokey Center for Life Sciences and Engineering, Technion, Haifa 320003, Israel (e-mail: yoavsh@bm.technion.ac.il).

The research was Funded by the European Union (ERC, DNAStorage, 865630). Views and opinions expressed are however those of the authors only and do not necessarily reflect those of the European Union or the European Research Council Executive Agency. Neither the European Union nor the granting authority can be held responsible for them. This work was also supported by the Gellman-Lasser Fund (11846) and the H2020 European Research Council Horizon
2020 (802567).}% 
}

\IEEEoverridecommandlockouts
\IEEEaftertitletext{}

\begin{document}

\maketitle

% ---- Abstract ---- %
\begin{abstract}
    \emph{DNA labeling} is a powerful tool in molecular biology and biotechnology that allows for the visualization, detection, and study of DNA at the molecular level. Under this paradigm, a DNA molecule is being \emph{labeled} by specific $k$ patterns and is then imaged. Then, the resulted image is modeled as a $(k+1)$-ary sequence in which any non-zero symbol indicates on the appearance of the corresponding label in the DNA molecule. The primary goal of this work is to study the \emph{labeling capacity}, which is defined as the maximal information rate that can be obtained using this labeling process. The labeling capacity is computed for almost any pattern of a single label and several results for multiple labels are provided as well. Moreover, we provide the optimal minimal number of labels of length one or two, over any alphabet of size $q$, that are needed in order to achieve the maximum labeling capacity of $\log_2(q)$. Lastly, we discuss the maximal labeling capacity that can be achieved using a certain amount of labels of length two.
\end{abstract}

% ---- Introduction ---- %
\section{Introduction}
\label{sec:introduction}

Labeling of DNA molecules with fluorescent markers is a widely used approach in molecular biology and medicine, with many applications in genomics and microbiology \cite{moter2000fluorescence, chen2018efficient, gruszka2021single}. This powerful tool allows for the visualization, detection, and study of DNA at the molecular level. Various techniques can be employed to achieve targeted labeling of DNA molecules, such as Fluorescence in situ hybridization (FISH) \cite{moter2000fluorescence}, CRISPR \cite{chen2018efficient,ma2015multicolor} and Methyltransferases \cite{deen2017methyltransferase}. Labeling is also done for other bio-molecules such as proteins and RNA, for applications in sensitive molecular analysis \cite{ohayon2019simulation, alfaro2021emerging} and studying gene expression and regulation \cite{young2020technical}, respectively.

DNA labeling is used for both specific target sequences and per-base labeling. Per-base labeling is used in DNA sequencing technologies that are based on sequencing by synthesis (such as illumina sequencing) \cite{CANARD19941}. Per-base labeling is also useful in Bisulphite sequencing to study methylation and epigenomics (genetic information beyond the genome sequence) \cite{mario2002}. Target sequence labeling is employed for species identification in clinical microbiology with FISH \cite{doi:10.3109/1040841X.2016.1169990}, studying DNA dynamics in living cells \cite{ma2015multicolor}, optical mapping~\cite{levy2013beyond, muller2017optical} (for genomic structural variation detection and species identification in microbiology), and the study of DNA-protein interactions, which are fundamental in understanding gene expression and regulation \cite{Dey2012}. By attaching a fluorescent label to DNA, researchers can visualize the interaction between DNA and proteins in real-time \cite{Dey2012}, providing insights into how DNA is packaged, organized, and interacts with proteins in the cell nucleus and how this affects gene expression. 

This work takes a first step towards mathematically modeling and analyzing the information rate that can be represented in labeled DNA molecules. More specifically, we study the \emph{labeling capacity}, which refers to the maximum information rate that can be stored by labeling DNA with specific sequence patterns as the labels.

In this work, the labeling process is formally modeled as follows. Assume the DNA sequence is given by $\bfx\in\{A,C,G,T\}^n$ and let $\bfalpha\in\{A,C,G,T\}^\ell$ be a short sequence which is being used as the \emph{label}. That is, the DNA sequence $\bfx$ is being labeled wherever $\bfalpha$ appears. As a result, a binary sequence $\bfz\in\{0,1\}^n$ is being received in which $z_i=1$ if and only if $(x_i,\ldots,x_{i+\ell-1})= \bfalpha$. For example, let $\bfalpha= \textcolor{red}{AC}$ be a label of length $\ell=2$. For $\bfx=AA\textcolor{red}{AC}GATG\textcolor{red}{ACAC}$, the received output binary sequence is ${\bfz=(0,0,\textcolor{red}1,0,0,0,0,0,\textcolor{red}1,0,\textcolor{red}1,0)}$. Clearly, there are other sequences, for example ${\bfy=TA\textcolor{red}{AC}TTTT\textcolor{red}{ACAC}\neq \bfx}$, which result with the same output binary sequence $\bfz$. So, the full capacity is not obtained and the goal of this work is to understand the maximum information rate of this paradigm, and the labeling capacity is referred as the asymptotic ratio between the number of information bits that can be stored and the length $n$. First, we show that the labeling capacity depends on the length of the used label. For example, for $|\bfalpha|=1$, any binary sequence can be achieved by the labeling process and thus the labeling capacity is 1. However, for length-2 labels $\bfalpha=(\alpha_1,\alpha_2)$, where $\alpha_1\neq \alpha_2$, binary sequences with two consecutive ones cannot be achieved, i.e., they satisfy the so-called $(d,k)$ \emph{run length limited} (\emph{RLL}) constraint~\cite{MRS2001} for $(d,k)=(1,\infty)$ and we show that the labeling capacity is the same as the capacity of the $(1,\infty)$ constraint. Besides the label's length, several more properties, such as its periodicity, may affect the labeling capacity. For example, the labeling capacity of $AA$ is larger than the one of the label $AC$ and we extend this result to find the labeling capacity of almost any label. The calculation of the labeling capacity for different labels is being used in order to organize all of the labels of lengths $\ell\leq5$ by their labeling capacity.

The labeling process can also be done using $k>1$ labels, for which any label is not a prefix of another label. In this case, the output is a sequence over $\{0,\ldots,k\}$. For example, let $\bfalpha_1= \textcolor{red}{AC}, \bfalpha_2= \textcolor{blue}{G}$ be two labels. For $\bfx=AA\textcolor{red}{AC}\textcolor{blue}{G}AT\textcolor{blue}{G}\textcolor{red}{ACAC}$ it holds that the received output sequence is $\bfz=(0,0,\textcolor{red}1,0,\textcolor{blue}2,0,0,\textcolor{blue}2,\textcolor{red}1,0,\textcolor{red}1,0)$. The definition of the labeling capacity is extended to multiple labels and we find this capacity when there is no overlap between the labels or for two non-cyclic labels in the special case where there is a unique way for the two labels to overlap each other. Another case that is being analyzed is the labeling capacity when using two labels that consist of a repetition of the same letter, of the same length, for example, using the labels $AAA,CCC$.

The following part of this work is dedicated to finding the minimal number of needed labels of a given length in order to obtain the maximum labeling capacity $\log_2(q)$, when $q$ is the alphabet size. For example, for labels of length~1, three different labels are necessary and sufficient to decode every sequence over $\{A,C,G,T\}^n$ and to have capacity 2. For labels of length 2, when $q=4$, it is clear that achieving labeling capacity 2 may be obtained using any 15 different labels of length two. Our main result here claims that this can be accomplished with 10 labels, and no less than 10 labels, i.e., 10 is the optimal number of labels to reach the maximum capacity. We also extend these results for arbitrary alphabet size $q$. Lastly, after finding the minimal number of labels of length two that are needed in order to obtain the maximal labeling capacity, we analyze the achievable labeling capacity in case of using
a smaller number of labels.

The rest of this paper is organized as follows. \Cref{sec:defs} formally defines the labeling channel and provides several useful definitions. \Cref{sec:cap} calculates the labeling capacity for a single label while considering its periodicity and overlap. At the end of \Cref{sec:cap}, all labels of lengths $\ell\leq5$ are being organized by their labeling capacity value. In~\Cref{sec:mul}, we extend the results for multiple labels. In~\Cref{sec:minimal}, the minimum number of labels needed to obtain the full capacity is studied for labels of length one or two. Lastly, in~\Cref{sec:largestCap}, the largest achievable capacity using two or nine labels of length two is being studied.

% ---- Background and Definitions---- %
\section{Definitions and Preliminaries}
\label{sec:defs}
Let $\Sigma_q$ denote the $q$-ary alphabet $\{0, 1,\ldots, q-1 \}$. For $q=4$, we mostly refer to the DNA alphabet, that is, ${\Sigma_4= \{A, C, G, T\}}$. Denote by $[n]$ the set $\{1, 2,\ldots, n\}$.
For a sequence ${\boldsymbol{x}}=(x_1,\ldots,x_n)\in\Sigma_q^n$, and $1\leq i \leq n-k+1$ , let ${\boldsymbol{x}}_{[i;k]}= (x_i,\ldots, x_{i+k-1})$. Moreover,
for $S\subseteq \Sigma^{n_1}_q$, $\bfu\in\Sigma^{n_2}_q$, let $S\circ \bfu \triangleq \{\bfw\in\Sigma^{n_1+n_2}_q| \exists \bfs\in S, \bfw=\bfs\bfu\}$.
A \emph{label} $\bfalpha\in\Sigma_q^\ell$ is a relatively short sequence over $\Sigma_q$. If the sequences and labels are over the DNA alphabet, we sometimes use strings instead of vectors for readibility (i.e., $AGGCGT\triangleq (A,G,G,C,G,T)$). Next, the \emph{labeling model} studied in this work is formally defined.
\begin{definition}
Let $\bfalpha_1,\ldots,\bfalpha_k$ be $k$ labels of lengths $\ell_1,\ldots,\ell_k$, respectively. Among the $k$ labels, there is no label that is a prefix of another label. Denote $\underline{\bfalpha}=(\bfalpha_1,\ldots,\bfalpha_k)$.
\begin{itemize}
    \item The $\underline{\bfalpha}$-\textbf{\emph{labeling sequence}} of  ${\boldsymbol{x}=(x_1,\ldots,x_n)\in\Sigma_q^n}$ is the sequence $L_{\underline{\bfalpha}}({\boldsymbol{x}}) = (c_1,\ldots,c_n)\in\Sigma_{k+1}^n$, in which $c_i=j$ if ${\boldsymbol{x}}_{[i;\ell_j]} = \bfalpha_j$ and $i\leq n-\ell_j+1$, and if such $j$ does not exist then $c_i=0$.
    \item A sequence ${\boldsymbol{u}}\in\{0,\ldots,k\}^n$ is called a \textbf{\emph{valid $\underline{\bfalpha}$-labeling sequence}} if there exists a sequence ${\boldsymbol{x}}\in\Sigma_q^n$ such that ${{\boldsymbol{u}}=L_{\underline{\bfalpha}}({\boldsymbol{x}})}$.
    \item Denote by $F_n(\underline{\bfalpha})$ the set of all valid $\underline{\bfalpha}$-labeling sequences of length $n$, which means, the image of the mapping  $L_{\underline{\bfalpha}}$. That is, $F_n(\underline{\bfalpha}) = \{L_{\underline{\bfalpha}}({\boldsymbol{x}}) | \bfx\in\Sigma_q^n\}$. 
    Denote the \textbf{\emph{labeling capacity}} of $\underline{\bfalpha}$ by $$\mathsf{cap}(\underline{\bfalpha}) \triangleq  \limsup_{n\to\infty}\frac{\log_2(|F_n(\underline{\bfalpha})|)}{n}.$$
\end{itemize}
\end{definition}
In case there is only a single label that is being used, another way to represent the labeling sequence is given in the next definition. Note that this definition is not well defined for multiple labels and thus we use it only for the single label case as it is merely used to ease the notations in some of the proofs throughout the paper.
\begin{definition}
    Let $\bfalpha$ be a label of length $\ell$.
    \begin{itemize}
        \item The \textbf{\emph{complete-$\bfalpha$-labeling sequence}} of ${\boldsymbol{x}}=(x_1,\ldots,x_n)\in\Sigma_q^n$ is the binary sequence $\widehat{L}_{\bfalpha}({\boldsymbol{x}}) = (c_1,\ldots,c_n)$, in which $\bfc_{[i;\ell]}=(1,\ldots,1)$ if ${\boldsymbol{x}}_{[i;\ell]} = \bfalpha$ and $i\leq n-\ell+1$.
        \item A sequence $\bfu\in\Sigma_2^n$ is called a \textbf{\emph{valid complete-$\bfalpha$-labeling sequence}} if there exists a sequence $\bfx\in\Sigma_q^n$ such that ${\bfu= \widehat{L}_\bfalpha(\bfx)}$.
        \item Denote by $\widehat{F}_n(\bfalpha)$ the set of all valid complete-$\bfalpha$-labeling sequences of length $n$.
    \end{itemize}
\end{definition}
Note that for any label $\bfalpha$ it holds that $|F_n(\bfalpha)|\geq |\widehat{F}_n(\bfalpha)|$ since there is a surjection from $F_n(\bfalpha)$ to $\widehat{F}_n(\bfalpha)$. For some labels, this function is also an injection. So, in order to compute the labeling capacity, for those labels, $\widehat{F}_n(\bfalpha)$ can be computed instead of $F_n(\bfalpha)$, for convenience.

\begin{example}
For $\bfalpha_1= \textcolor{red}{CG},  \bfalpha_2= \textcolor{blue}A$, ${\boldsymbol{x}}=\textcolor{blue}AC\textcolor{red}{CG}G\textcolor{red}{CG}\textcolor{blue}AT\textcolor{blue}A$, it holds that $L_{(\bfalpha_1,\bfalpha_2)}({\boldsymbol{x}})=(\textcolor{blue}2,0,\textcolor{red}1,0,0,\textcolor{red}1,0,\textcolor{blue}2,0,\textcolor{blue}2)$. 
Moreover, $\widehat{L}_{\bfalpha_1}(\bfx)=(0,0,\textcolor{red}1,\textcolor{red}1,0,\textcolor{red}1,\textcolor{red}1,0,0,0)$ and $L_{\bfalpha_1}(\bfx)=(0,0,\textcolor{red}1,0,0,\textcolor{red}1,0,0,0,0)$.
\end{example}
The following definitions will be helpful in order to discuss different types of labels.
\begin{definition}
Let $\bfalpha, \bfalpha'$ be a label of length $\ell, \ell'$, respectively.
\begin{itemize}
    \item The \textbf{\emph{period}} of $\bfalpha$ is $\cP(\bfalpha)\triangleq \min\{p\in[\ell] : p|\ell, \bfalpha_{[1;p]}= \bfalpha_{[(t-1)p+1;p]}$ for $t\in[\frac{\ell}{p}]\}$. In case $p=\ell$, there is no period in $\bfalpha$ and the label is called a \textbf{\emph{non-periodic}} label.
    \item The \textbf{\emph{overlap}} between $\bfalpha$ and $\bfalpha'\neq\bfalpha$ is $\cO(\bfalpha,\bfalpha')\triangleq\max\{r\in[\min\{\ell,\ell'\}]: \bfalpha'_{[1;r]} = \bfalpha_{[\ell'-r+1;r]}\}$. In other words, $\cO(\bfalpha,\bfalpha')$ is the maximal size of a suffix of $\bfalpha$ which is identical to a prefix of $\bfalpha'$. In case $\cO(\bfalpha,\bfalpha')$ does not exist, we define $\cO(\bfalpha,\bfalpha')\triangleq 0$.
    The labels $\bfalpha$ and $\bfalpha'$ are called \textbf{\emph{overlapping labels}} if $\cO(\bfalpha,\bfalpha')>0$ or $\cO(\bfalpha',\bfalpha)>0$.
    \item The \textbf{\emph{cyclic overlap}} of $\bfalpha$ is $\cO(\bfalpha)\triangleq \cO(\bfalpha_{[1;\ell-1]},\bfalpha_{[2;\ell-1]})$ if $\ell>1$, and otherwise $\cO(\bfalpha)\triangleq0$. In case 
    $\cO(\bfalpha)= 0$,
    $\bfalpha$ is called a \textbf{\emph{non-cyclic label}}.
\end{itemize}
\end{definition}
Note that a periodic label is also a cyclic label but a cyclic label is not necessarily periodic. For a periodic label $\bfalpha$ of length $\ell$, it holds that $\cO(\bfalpha)=\ell- \cP(\bfalpha)$.
The next example exemplifies the definitions above. 
\begin{example}
The labels $\bfalpha_1=CGCGCG$ and $\bfalpha_2=GATG$ are overlapping labels. It holds that $\cO(\bfalpha_1,\bfalpha_2)=1$ and $\cO(\bfalpha_2,\bfalpha_1)=0$. Moreover, $\cO(\bfalpha_1)=4$ and $\cP(\bfalpha_1)=2$. In contrast, $\cO(\bfalpha_2)=1$ but it is a non-periodic label.
\end{example}

One of the goals in this work is to calculate the labeling capacity using one or more labels. Some of the results will be derived by drawing a connection to constrained systems. In order to establish this connection, several more definitions are introduced as described in~\cite{MRS2001}.
\begin{definition}
A \textbf{\emph{finite labeled\footnote{Contrary to the definition of labeling in this work, here the meaning of labeling refers to giving labels to the edges of the graph.} directed graph}} $G = (V,E,L)$ is a graph that consists of a finite set of states $V$, a finite set of edges $E$, and an edge labeling $L: E \xrightarrow{} \Sigma_q$.
A sequence $\bfw$ over $\Sigma_q$ is \textbf{\emph{generated}} by $\pi$ (and $G$) if $\pi$ is a path in $G$ which is labeled by the sequence $\bfw$. A labeled graph $G$ is \textbf{\emph{deterministic}} if the outgoing edges from each state are labeled distinctly.
A \textbf{\emph{constraint}} $S$ is the set of all sequences over $\Sigma_q$ that are generated by a labeled graph $G$. In this case, it is said that $G$ \textbf{\emph{presents}} $S$ and it is denoted by $S=S(G)$.
Denote the set of sequences of length $n$ in the constraint $S$ by $S(n)=|S\cap\Sigma_q^n|$.
It is known that for each constraint, there exists a deterministic graph that presents it. The \textbf{\emph{capacity}} of a constraint $S$
is $$\mathsf{cap}(S) \triangleq  \limsup_{n\to\infty}\frac{1}{n}\log_2(S(n)).$$ 
For a deterministic presentation $G$ of $S$, it holds that $\mathsf{cap}(S) =\log_2(\lambda(A_G))$, where $\lambda(A_G)$ is the spectral
radius (also known as Perron eigenvalue), which is the largest real eigenvalue in absolute value of the adjacency matrix of $G$. 
\end{definition}

Some of the results in the paper will be connected to a specific constraint, known as the the run-length limited (RLL) constraint, as described in the next definition.
\begin{definition}
A sequence over $\Sigma_q$ satisfies the \textbf{\emph{$(q,d,k)$-RLL constraint}} if between every two consecutive non-zero symbols there are at least $d$ zeroes and there is no run of zeroes of length $k+1$. Denote the set of all sequences of length $n$ that satisfy the $(q,d,k)$-RLL constraint by $\cC_{q,d,k}(n)$. For the case of $q=2$, this constraint is called the \textbf{\emph{$(d,k)$-RLL constraint}}, and $\cC_{2,d,k}$ will be denoted by $\cC_{d,k}$.
\end{definition}
It has been proven that $\mathsf{cap}(\cC_{q,d,\infty})= \log_2\lambda$ when $\lambda$ is the largest real root of $x^{d+1}-x^{d}-(q-1)$~\cite{McLaughlin95}.

In this work, we study the following two problems.
 \begin{problem}\label{prob1}
    The $k$-labeling capacity problem:
    Let $\underline{\bfalpha}=(\bfalpha_1,\ldots,\bfalpha_k)$ be $k$ labels. Find the labeling capacity $\mathsf{cap}(\underline{\bfalpha})$.
\end{problem}
\begin{problem}\label{prob2}
     Find the minimal $s$ such that there exists $\underline{\bfalpha}=(\bfalpha_1,\ldots,\bfalpha_s)$ of $s$ labels, each of length $\ell\geq 1$, and ``almost" every $\boldsymbol{x}\in\Sigma_q^n$ can be determined given its $\underline{\bfalpha}$-labeling sequence. More specifically, we are interested in computing the minimal number of labels of a fixed length $\ell$ that are needed in order to gain capacity of $\log_2q$. Mathematically, the problem is to find the value of
     $$s(\ell,q)\triangleq \min \{ s\in\mathbb{N}| \exists \underline{\bfalpha}\in\Sigma^s_{q^\ell} ,\mathsf{cap}(\underline{\bfalpha})=\log_2q \}.$$
\end{problem}

\section{The Labeling Capacity of a Single Label}\label{sec:cap}
In this section, we provide a full solution to the labeling capacity in case a single label is used. More specifically, we solve Problem~\ref{prob1} in case $k=1$ for different types of labels, such as non-cyclic and periodic labels. Building upon our extensive analysis, at the end of this section, we also show how to order all of the labels of length $\ell\leq5$ according to their labeling capacity.
\subsection{Non-Cyclic Labels}
This section is focused on the case of using a non-cyclic label. The next theorem provides the labeling capacity in this case. The following corollary shows that for longer non-cyclic labels, the labeling capacity is smaller. 
\begin{theorem}\label{non-cyclic}
Let $\bfalpha\in\Sigma_q^\ell$ be a non-cyclic label of length $\ell$. Then, $\mathsf{cap}(\bfalpha)=\mathsf{cap}(\cC_{{\ell-1},\infty})$. That is, $\mathsf{cap}(\bfalpha)=\log_2\lambda$ when $\lambda$ is the largest real root of $x^\ell-x^{\ell-1}-1$.
\end{theorem}
\begin{IEEEproof}
Let $\bfalpha\in\Sigma_q^\ell$ be a non-cyclic label of length $\ell$. For $\boldsymbol{x}\in\Sigma_q^n$, let $\boldsymbol{y}\in\Sigma_2^n$ be the $\bfalpha$-labeling sequence of $\boldsymbol{x}$, i.e., $\bfy = L_{\bfalpha}(\bfx)$. By definition, it holds that $y_i = 1$ if and only if ${\boldsymbol{x}}_{[i;\ell]} = \bfalpha$. Since the label $\bfalpha$ is non-cyclic it holds that if ${\boldsymbol{x}}_{[i;\ell]} = \bfalpha$, then ${\boldsymbol{x}}_{[j;\ell]} \neq \bfalpha$ for $i+1\leq j \leq i+\ell-1$. Hence, for every $\bfalpha$-labeling sequence it holds that after each one there are at least $\ell-1$ zeroes, i.e., ${\boldsymbol{y}}$  satisfies the $(\ell-1,\infty)$-RLL constraint and ends with $\ell-1$ zeroes. So, $F_n(\bfalpha)\subseteq \cC_{\ell-1,\infty}(n-(\ell-1))\circ0^{\ell-1}$. In order to prove inclusion in the other direction, let $\bfu\in\cC_{\ell-1,\infty}(n-\ell+1)\circ0^{\ell-1}$. Let $\bfv\in\Sigma_q^n$ be such that $\bfv_{[i;\ell]}= \bfalpha$ if and only if $u_i=1$. It holds that $L_\bfalpha(\bfv)=\bfu$. From the definition of ${\cC_{\ell-1,\infty}(n-(\ell-1))\circ0^{\ell-1}}$, after each one in $\bfu$ there are at least $\ell-1$ zeroes and $|\bfalpha|=\ell$, so such a $\bfv$ exists.
Hence, for 
$n\geq\ell-1$, $|F_n(\bfalpha)|= |\cC_{\ell-1,\infty}(n-\ell+1)\circ0^{\ell-1}|$.
So,
\begin{align*}
 \mathsf{cap}(\bfalpha)& =\limsup\limits_{n\to\infty}\frac{\log_2(|F_n(\bfalpha)|)}{n}   \\
& =\limsup\limits_{n\to\infty}\frac{\log_2(|\cC_{\ell-1,\infty}(n-\ell+1)|)\cdot(n-\ell+1)}{n\cdot(n-\ell+1)} \\
& =\limsup\limits_{n\to\infty}\frac{\mathsf{cap}(\cC_{\ell-1,\infty})\cdot(n-\ell+1)}{n} \\
& =\mathsf{cap}(\cC_{\ell-1,\infty}) = \log_2\lambda,
\end{align*}
where $\lambda$ is the largest real root of $x^\ell-x^{\ell-1}-1$.
\end{IEEEproof}

The results in this section will show that the labeling capacity for a specific label is the base two logarithm of the largest real root of some polynomial. It means that each of the polynomials that will be analyzed has a root which is at least one. Otherwise, the labeling capacity value would be negative, which leads to a contradiction. Moreover, for the polynomials that will be shown, one is not a root. Thus, each of those polynomials has a root which is strictly greater than one.

The next corollary follows directly from~\Cref{non-cyclic} and shows the connection between the capacities of non-cyclic labels of different lengths.
\begin{corollary}\label{Different length non cyclic}
    Let $\ell>0$ and let $\bfalpha_1,\bfalpha_2$ be a non-cyclic label of length $\ell,\ell+k$ for $k\in\N$, respectively. It holds that $\mathsf{cap}(\bfalpha_1)> \mathsf{cap}(\bfalpha_2)$. In other words, for non-cyclic
    labels, the shorter the label is, the larger the labeling capacity is.
\end{corollary}
\begin{IEEEproof}
    Let $\ell> 0$ and $\bfalpha_1,\bfalpha_2$ be a non-cyclic label of length $\ell,\ell+k$, respectively. From~\Cref{non-cyclic}, $\mathsf{cap}(\bfalpha_1)=\log_2\lambda_1,\mathsf{cap}(\bfalpha_2)=\log_2\lambda_2$, when $\lambda_1,\lambda_2$ is the largest real root of the polynomial $x^\ell-x^{\ell-1}-1,x^{\ell+k}-x^{\ell+k-1}-1$, respectively. 
    As mentioned earlier, it holds that $\lambda_1,\lambda_2>1$. From the equation $\lambda_1^{\ell}-\lambda_1^{\ell-1}-1=0$, it holds that $\lambda_1^{\ell}-\lambda_1^{\ell-1}=\lambda_1^{\ell-1}\cdot(\lambda_1-1)=1$. From the equation $\lambda_2^{\ell+k}-\lambda_2^{\ell+k-1}-1=0$, it holds that $\lambda_2^{\ell-1}\cdot(\lambda_2-1)=\frac{1}{\lambda_2^k}<1$. The function $f(x)=x^{\ell-1}\cdot(x-1)$ is an increasing function for $x>1$. So, from the equations above, it holds that $\lambda_1>\lambda_2$ and so, $\mathsf{cap}(\bfalpha_1)=\log_2(\lambda_1)>\mathsf{cap}(\bfalpha_2)=\log_2(\lambda_2)$.
\end{IEEEproof}

\subsection{Labels with Non-Cyclic Period}
This section analyzes the case of using periodic labels, where the period of the label is non-cyclic. Before we continue with the labeling capacity for this case, the next example motivates the solution that will be presented afterwards.

\begin{example}
\label{periodicExample}
Let $\bfalpha=CGCG$, so $\cP(\bfalpha)=2$. For this label, it holds that there is a bijection between $F_n(\bfalpha)$ and $\widehat{F}_n(\bfalpha)$. So, in order to compute the labeling capacity, $\widehat{F}_n(\bfalpha)$ will be computed. Let $\boldsymbol{x}\in\{A,C,G,T\}^n$ and let $\boldsymbol{y}\in\Sigma_2^n$ be the complete-$\bfalpha$-labeling sequence of $\boldsymbol{x}$, i.e., $\bfy = \widehat{L}_{\bfalpha}(\bfx)$. It holds that if $\bfx_{[i;4]} = CGCG$, then $\bfy_{[i,4]} = (1,1,1,1)$. So, if for $k\geq2$, $\bfx_{[i;2k]}$ consists of a run of $k$ $CG$s and $\bfx_{[i+2k;2]}\neq CG$, then $\bfy_{[i;2k]} = (1,1,\ldots,1)$ and $y_{i+2k}=0$, because otherwise this implies that $\bfx_{[i+2k-2;4]}= CGCG$ but $\bfx_{[i+2k;2]} \neq CG$. It can be concluded that every valid complete-$\bfalpha$-labeling sequence is a binary sequence in which the length of every run of ones is even and is at least four. Denote this set by $S_{E\geq4}$ and $S_{E\geq4}(n)\triangleq S_{E\geq4}\cap\Sigma_2^n$. After proving that $\widehat{F}_n(\bfalpha)\subseteq S_{E\geq4}(n)$, in order to prove equality between those sets, we prove inclusion in the other direction next. Let $\bfu\in S_{E\geq4}(n)$ and let $\bfv\in\Sigma_4^n$ be such that for $k\geq2$, $\bfv_{[i;2k]}=CGCG\cdots CG$ if $\bfu_{[i;2k]}=(1,\ldots,1)$ and $u_{i-1}=0$ or $i=0$. It holds that $\widehat{L}_{\bfalpha}(\bfv)=\bfu$.

Hence, we have that  $|\widehat{F}_n(\bfalpha)|=|S_{E\geq4}(n)|$. Denote the constraint that is presented in the graph in~\Cref{periodicExampleGraph} by $S$.
\begin{figure}[h]
\centering
    \begin{tikzpicture}[->,>=stealth',shorten >=1pt,thick,round/.style={draw,circle,thick,minimum size=8mm,thick},]
\SetGraphUnit{3} 
\node[round] at (0,0) (A) [] {$v_0$};
\node[round] (B) at (1.65,0) [] {$v_1$};
\node[round] (C) at (3.3,0) [] {$v_2$};
\node[round] (D) at (4.95,0) [] {$v_3$};
\node[round] (E) at (6.6,0) [] {$v_4$};

\Loop[dist=1cm,dir=NO,label=$0$,labelstyle=above](A)
\draw[->] (A) to [right] node [above] {1} (B);
\draw[->] (B) to [right] node [above] {1} (C);
\draw[->] (C) to [right] node [above] {1} (D);
\draw[->] (D) to [right] node [above] {1} (E);
\draw[->] (E) to [bend left] node [below] {0} (A);
\draw[->] (E) to [bend right] node [above] {1} (D);
\end{tikzpicture}
\caption{Graph presentation of the constraint in~\Cref{periodicExample}.}
\label{periodicExampleGraph}
\end{figure}
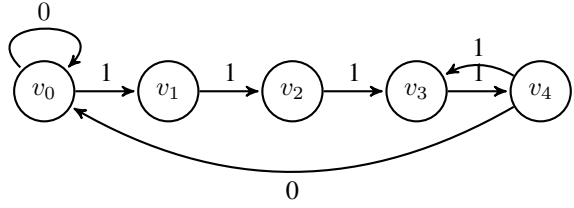
From the structure of the graph, for any sequence $\bfy'$ of length $n-6$, when $n\geq 6$, that is generated by the graph, there exist sequences ${\bfy'',\bfy'''\in\Sigma^3_2}$, such that ${\bfy\triangleq\bfy'''\circ\bfy'\circ\bfy''\in S_{E\geq4}(n)\subseteq S(n)}$.
Hence, we have an injection from $S(n-6)$ to $S_{E\geq4}(n)$ and ${|S(n-6)|\leq|S_{E\geq4}(n)|\leq|S(n)|}$. Hence, $\mathsf{cap}(\underline{\bfalpha})= \mathsf{cap}(S)$. The adjacency matrix of this graph is

\begin{small}
$$\begin{pmatrix}
1 & 1 & 0 & 0 & 0\\
0 & 0 & 1 & 0 & 0\\
0 & 0 & 0 & 1 & 0\\
0 & 0 & 0 & 0 & 1\\
1 & 0 & 0 & 1 & 0\\
\end{pmatrix},$$
\end{small}
and its characteristic polynomial is $x^5-x^4-x^3+x^2-1$. Thus, the capacity of $S$ is $\log_2(\lambda)$ when $\lambda\approx1.44$ is the largest real root of this polynomial. 
\end{example}

This example leads to the general case of calculating the labeling capacity of a label with period $p<\ell$ which is non-cyclic. 
\begin{theorem}
\label{periodicTheorem}
Let $\bfalpha$ be a label of length $\ell$ with a non-cyclic period $\cP(\bfalpha)=p<\ell$. It holds that $\mathsf{cap}(\bfalpha)=\log_2(\lambda)$ when $\lambda$ is the largest real root of the polynomial $x^{\ell+1}-x^{\ell}-x^{\ell-p+1}+x^{\ell-p}-1$.
\end{theorem}
\begin{IEEEproof}
Let $\bfalpha$ be a label of length $\ell$ with a non-cyclic period $p=\cP(\bfalpha)<\ell$. Since in this case there is a bijection between $F_n(\bfalpha)$ and $\widehat{F}_n(\bfalpha)$, $\widehat{F}_n(\bfalpha)$ will be computed in order to find the labeling capacity. Denote $\bfalpha'=\bfalpha_{[1;p]}$.
Let $\boldsymbol{x}\in\Sigma_q^n$ and let $\bfy = \widehat{L}_{\bfalpha}(\bfx)$. It holds that if $\bfx_{[i;\ell]} = \bfalpha$, then $\bfy_{[i;\ell]} = (1,\ldots,1)$. So, if there exists $k\geq\frac{\ell}{p}$ such that $\bfx_{[i;pk]}$ consists of the concatenation of $\bfalpha'$ $k$ times and $\bfx_{[i+pk;p]}\neq \bfalpha'$, then $\bfy_{[i;pk]}=(1,\ldots,1)$ and $y_{i+pk}=0$, since otherwise this implies that $\bfx_{[i+pk-(\ell-p);\ell]}= \bfalpha$ but $\bfx_{[i+pk;p]} \neq \bfalpha'$. As a result, every valid complete-$\bfalpha$-labeling sequence is a binary sequence in which the length of every run of ones is at least $\ell$ and is divisible by $p$. Denote this set by $S_{p\geq\ell}$ and $S_{p\geq\ell}(n)\triangleq S_{p\geq\ell}\cap\Sigma_2^n$. Next, we show that $S_{p\geq\ell}(n)\subseteq\widehat{F}_n(\bfalpha)$. Let $\bfu\in S_{p\geq\ell}(n)$ and let $\bfv\in\Sigma_q^n$ be a sequence in which if for $k\geq\frac{\ell}{p}$, $\bfu_{[i,pk]}=(1,\ldots,1)$ and $u_{i-1}=0$ or $i=0$, then $\bfv_{[i,pk]}=\bfalpha'\circ\cdots\circ\bfalpha'$. It holds that $\widehat{L}_\bfalpha(\bfv)=\bfu$.
So, $|\widehat{F}_n(\bfalpha)|=|S_{p\geq\ell}(n)|$. Denote the constraint that is presented in~\Cref{periodicTheoremGraph} by $S$, where the missing edges of the graph are labeled with $1$.

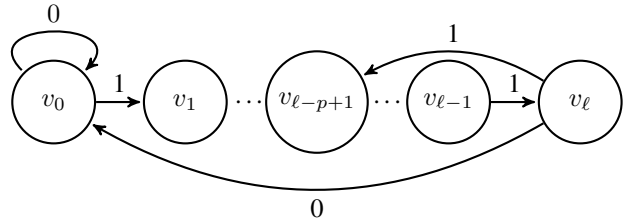
\begin{figure}[h]
\centering
    \begin{tikzpicture}[->,>=stealth',shorten >=1pt,thick,round/.style={draw,circle,thick,minimum size=11mm,thick},]
\SetGraphUnit{2} 
\node[round] at (0,0) (A) [] {$v_0$};
\node[round] (B) at (1.75,0) [] {$v_1$};
%\node[round] (C) at (2.8,0) [] {$v_2$};
\node[round] (D) at (3.5,0) [] {$v_{\ell-p+1}$};
\node[round] (E) at (5.25,0) [] {$v_{\ell-1}$};
\node[round] (F) at (7,0) [] {$v_{\ell}$};

\Loop[dist=1cm,dir=NO,label=$0$,labelstyle=above](A)
\draw[->] (A) to [right] node [above] {1} (B);
\draw[->] (F) to [bend right] node [above] {1} (D);
\node at ($(B)!.5!(D)$) {\ldots};
\node at ($(D)!.55!(E)$) {\ldots};
\draw[->] (E) to [right] node [above] {1} (F);
\draw[->] (F) to [bend left] node [below] {0} (A);
\end{tikzpicture}
\caption{Graph presentation of the constraint in~\Cref{periodicTheorem}.}
\label{periodicTheoremGraph}
\end{figure}

From the structure of the graph, for any sequence $\bfy'$ of length $n-2(\ell-1)$, when $n\geq 2(\ell-1)$, that is generated by the graph, there exist sequences ${\bfy'',\bfy'''\in\Sigma^{\ell-1}_2}$, such that ${\bfy\triangleq\bfy'''\circ\bfy'\circ\bfy''\in S_{p\geq\ell}(n)\subseteq S(n)}$. So, we have an injection from $S(n-2(\ell-1))$ to $S_{p\geq\ell}(n)$ and ${|S(n-2(\ell-1))|\leq|S_{p\geq\ell}(n)|\leq|S(n)|}$. It implies that $\mathsf{cap}(\underline{\bfalpha})= \mathsf{cap}(S)$. The adjacency matrix of this graph is

\begin{small}
$$A=\begin{pmatrix}
1 & 1 & 0 & 0 & \cdots & 0\\
0 & 0 & 1 & 0 & \cdots & 0\\
0 & 0 & 0 & 1 & \cdots & 0\\
\vdots & \vdots & \vdots & \vdots & \ddots & \vdots\\
0 & 0 & 0 & 0 & 0 & 1\\
1 & 0 & 0 & 1 & \cdots & 0\\
\end{pmatrix},$$
\end{small}
where $A_{i,j}=1$ for $(i,j)$ such that: $i=0, j=0$ or $i=j-1$ or $i=\ell, j=0$ or $i=\ell, j=\ell-p+1$ and otherwise, $A_{i,j}=0$.
It can be shown that the characteristic polynomial of this matrix is $x^{\ell+1}-x^{\ell}+x^{\ell-p}-x^{\ell-p+1}-1$. Thus, $\mathsf{cap}(S)=\log_2(\lambda)$ where $\lambda$ is the largest real root of the last polynomial. 
\end{IEEEproof}

The following two corollaries show the order, with respect to labeling capacity values, between periodic labels with non-cyclic period. The first corollary discusses this type of labels of the same length with different sizes of period, while the second one discusses this type of labels of different lengths and the same period value. 

\begin{corollary}\label{Ordering periodic}
    Let $\ell>0$ and let $\bfalpha_1,\bfalpha_2$ be periodic labels of length $\ell$ with non-cyclic period, and, $\cP(\bfalpha_1)<\cP(\bfalpha_2)$. Then, it holds that $\mathsf{cap}(\bfalpha_1)> \mathsf{cap}(\bfalpha_2)$. In other words, for periodic labels with non-cyclic period of the same length, the smaller the period value is, the larger the labeling capacity is.
\end{corollary}

\begin{IEEEproof}
    Let $\bfalpha_1$ be a periodic label of length $\ell$ with $\cP(\bfalpha_1)=p<\ell$. From~\Cref{periodicTheorem}, it holds that $\mathsf{cap}(\bfalpha_1)=\log_2(\lambda_1)$ when $\lambda_1$ is the largest real root of the polynomial $x^{\ell+1}-x^{\ell}-x^{\ell-p+1}+x^{\ell-p}-1$. Assume that there is another periodic label $\bfalpha_2$ of length $\ell$ with $\cP(\bfalpha_2)=p+k <\ell, k\in\N$. It holds that $\mathsf{cap}(\bfalpha_2)=\log_2(\lambda_2)$ when $\lambda_2$ is the largest real root of the polynomial $x^{\ell+1}-x^{\ell}-x^{\ell-p-k+1}+x^{\ell-p-k}-1$. As mentioned earlier, it holds that $\lambda_1,\lambda_2>1$. From the equation $\lambda_1^{\ell+1}-\lambda_1^{\ell}-\lambda_1^{\ell-p+1}+\lambda_1^{\ell-p}-1=0$, it holds that $\frac{\lambda_1^\ell(\lambda_1-1)-1}{\lambda_1^{\ell-p}(\lambda_1-1)}=1$. From the equation $\lambda_2^{\ell+1}-\lambda_2^{\ell}-\lambda_2^{\ell-p-k+1}+\lambda_2^{\ell-p-k}-1=0$, it holds that $\frac{\lambda_2^\ell(\lambda_2-1)-1}{\lambda_2^{\ell-p}(\lambda_2-1)}=\frac{1}{\lambda_2^k}<1$. The function $f(x)=\frac{x^\ell-x^{\ell-1}-1}{x^{\ell-p}(x-1)}$ is an increasing function for $x>1$. So, it holds that $\lambda_1>\lambda_2$ and $\mathsf{cap}(\bfalpha_1)=\log_2(\lambda_1)>\log_2(\lambda_2) =\mathsf{cap}(\bfalpha_2)$.
\end{IEEEproof}

\begin{corollary}\label{Different length same period}
    Let $\ell>0$ and let $\bfalpha_1,\bfalpha_2$ be two periodic labels with non-cyclic periods with the same period size $\cP(\bfalpha_1)=\cP(\bfalpha_2)=p$ of lengths $\ell,\ell+k$ when $k\in \N$, respectively. Then, it holds that $\mathsf{cap}(\bfalpha_1)> \mathsf{cap}(\bfalpha_2)$.
\end{corollary}

\begin{IEEEproof}
    Let $\ell> 0$ and $\bfalpha_1$ be a periodic label of length $\ell$, $\cP(\bfalpha_1)=p<\ell$. From~\Cref{periodicTheorem}, $\mathsf{cap}(\bfalpha_1)=\log_2\lambda_1$ when $\lambda_1$ is the largest real root of $p(x)= x^{\ell+1}-x^{\ell}-x^{\ell-p+1}+x^{\ell-p}-1$. Let $\bfalpha_2$ be a periodic label of length $\ell+k$ for $k\in\N$, $\cP(\bfalpha_2)=p$. From~\Cref{periodicTheorem}, $\mathsf{cap}(\bfalpha_2)=\log_2\lambda_2$ when $\lambda_2$ is the largest real root of $x^{\ell+k+1}-x^{\ell+k}-x^{\ell+k-p+1}+x^{\ell+k-p}-1$. As mentioned earlier, it holds that $\lambda_1,\lambda_2>1$. From the equation $\lambda_1^{\ell+1}-\lambda_1^{\ell}-\lambda_1^{\ell-p+1}+\lambda_1^{\ell-p}-1=0$, it holds that $\lambda_1^{\ell-p}\cdot(\lambda_1^p-1)\cdot(\lambda_1-1)=1$. From the equation $\lambda_2^{\ell+k+1}-\lambda_2^{\ell+k}-\lambda_2^{\ell+k-p+1}+\lambda_2^{\ell+k-p}-1=0$, it holds that $\lambda_2^{\ell-p}\cdot(\lambda_2^p-1)\cdot(\lambda_2-1)=\frac{1}{\lambda_2^k}<1$. The function $f(x)=x^{\ell-p}\cdot(x^p-1)\cdot(x-1)$ is an increasing function for $x>1$. So, from the equations above, it holds that $\lambda_1>\lambda_2$ and so, $\mathsf{cap}(\bfalpha_1)=\log_2(\lambda_1)>\mathsf{cap}(\bfalpha_2)=\log_2(\lambda_2)$.
\end{IEEEproof}

\subsection{Non-Periodic Labels with Cyclic-Overlap}
This section studies the case of using a label with cyclic overlap that holds some properties.
In order to present the theorem regarding the capacity of labels with cyclic overlap, a new definition will be given first.
\begin{definition}\label{almostPeriodicDef}
    A label $\bfalpha$ of length $\ell$ is called an \textbf{\emph{almost-periodic}} label if there exists a sub-label of $\bfalpha$, that will be denoted by $\bfalpha'$, of length $p'$ and an integer $t<p'$ such that $\bfalpha=\bfalpha'\circ\bfalpha'\circ\cdots\circ\bfalpha'\circ\bfalpha'_{[1;t]}$. The sub-label $\bfalpha'$ is called the \textbf{\emph{partial-period}} of $\bfalpha$ and $\bfalpha'_{[1;t]}$ is called the \textbf{\emph{almost-periodic suffix}} of $\bfalpha$.
\end{definition}
Note that an almost periodic label is also a cyclic label.

\begin{example}
The labels $\bfalpha_1= ACACA$, $\bfalpha_2= CGTCGTC$ are almost periodic labels. The partial-period of $\bfalpha_1$ is $AC$ and the partial-period of $\bfalpha_2$ is $CGT$. Both of these labels have cyclic overlap and $\cO(\bfalpha_1)=3$, $\cO(\bfalpha_2)=4$.  
\end{example}

Next, we consider the case of a non-periodic label of length $\ell$ with a cyclic overlap $r>0$. In particular, in the following example and theorem we study this case, assuming at least one of the following conditions holds.
 \begin{enumerate}
     \item The cyclic overlap does not contain a cyclic overlap. For example, the label $ACGAC$ holds this condition because $AC$ does not have a cyclic overlap. However, $ACAGACA$ does not hold this condition.
     \item The label is an almost periodic label and the almost-periodic suffix of the label does not contain a cyclic overlap. For example, the label $ACGACGAC$ holds this condition because $AC$ does not have a cyclic overlap. However, $AAGAAGAA$ does not hold this condition since $AA$ is cyclic.
 \end{enumerate}
 The first condition implies that there is only one way in which the label can overlap with itself. The second condition forces similar behavior.

\begin{example}
\label{CyclicOverlapExample}
Let $\bfalpha= ATA$, so the $\cO(\bfalpha)=1$ and $\cP(\bfalpha)=3$. Let $\boldsymbol{x}\in\{A,C,G,T\}^n$ and let $\boldsymbol{y}\in\Sigma_2^n$ be the $\bfalpha$-labeling sequence of $\boldsymbol{x}$, i.e., $\bfy = L_{\bfalpha}(\bfx)$. From the definition of $\bfalpha$-labeling sequences, it holds that if $\bfx_{[i;3]} = ATA$, then $y_{i} = 1$. So, if $\bfx_{[i;2k]}$ consists of a repetition of the sub-label $AT$ $k\geq1$ times, $x_{i+2k}= A$, and $\bfx_{[i+2k+1;2]}\neq TA$, then $\bfy_{[i;2k+1]}$ is of the form $(10)^{k-1}100$. Note that in contrarily to the case of a periodic label, $y_{i+2k+1}$ can be one if the label is concatenated to itself. It can be concluded that every valid $\bfalpha$-labeling sequence is a binary sequence of the form $(0+(10)^*(100))^*$. Denote this set by $S_r$ and $S_r(n)\triangleq S_r\cap\Sigma_2^n$. Next, we show that $S_r(n)\subseteq F_n(\bfalpha)$. Let $\bfu\in\Sigma_2^n$ be a binary sequence in $S_r(n)$.
Let $\bfv\in\Sigma_4^n$ be a sequence in which if $u_i= 1$, then $\bfv_{[i;3]}=ATA$. It holds that $L_{\bfalpha}(\bfv)= \bfu$.
As a result, it holds that $|F_n(\bfalpha)|=|S_r(n)|$. Denote the constraint that is presented in~\Cref{ATAgraph} by $S$.

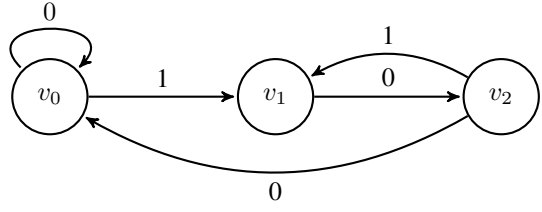
\begin{figure}[h]
\centering
    \begin{tikzpicture}[->,>=stealth',shorten >=1pt,thick,round/.style={draw,circle,thick,minimum size=1cm,thick},]
\SetGraphUnit{3} 
\node[round] at (0,0) (A) [] {$v_0$};
\node[round] (B) at (3,0) [] {$v_1$};
\node[round] (C) at (6,0) [] {$v_2$};

\Loop[dist=1cm,dir=NO,label=$0$,labelstyle=above](A)
\draw[->] (A) to [right] node [above] {1} (B);
\draw[->] (B) to [right] node [above] {0} (C);
\draw[->] (C) to [bend right] node [above] {1} (B);
\draw[->] (C) to [bend left] node [below] {0} (A);
\end{tikzpicture} 
\caption{Graph presentation of the constraint in~\Cref{CyclicOverlapExample}.}
\label{ATAgraph}

\end{figure}

From the structure of the graph, for any sequence $\bfy'$ of length $n-2$, when $n\geq 2$, that is generated by the graph, it holds that ${\bfy\triangleq\bfy'\circ00\in S_r(n)\subseteq S(n)}$.
Hence, we have an injection from $S(n-2)$ to $S_r(n)$ and ${|S(n-2)|\leq|S_r(n)|\leq|S(n)|}$. Hence, $\mathsf{cap}(\bfalpha)= \mathsf{cap}(S)$. The adjacency matrix of this graph is
$$\begin{pmatrix}
1 & 1 & 0\\
0 & 0 & 1\\
1 & 1 & 0\\
\end{pmatrix},$$
and the characteristic polynomial of this matrix is $x^3-x^2-x$. Thus, the capacity of $S$ is $\log_2(\lambda)$ when $\lambda\approx 1.618$ is
the largest real root of this polynomial. 
\end{example}

\begin{theorem}
\label{CyclicOverlapTheorem} 
Let $\bfalpha$ be a non-periodic label of length $\ell$ with a cyclic overlap $\cO(\bfalpha)=r>0$. Moreover, one of the following conditions holds:
\begin{enumerate}
    \item The cyclic overlap does not contain a cyclic overlap.
    \item The label is an almost periodic label and the almost-periodic suffix of the label does not contain a cyclic overlap.
 \end{enumerate}
It holds that $\mathsf{cap}(\bfalpha)=\log_2(\lambda)$ when $\lambda$ is the largest real root of the polynomial $x^{\ell}-x^{\ell-1}-x^{r}+x^{r-1}-1$. 
\end{theorem}

\begin{IEEEproof}
Let $\bfalpha$ be a non-periodic label of length $\ell$ with a cyclic overlap $\cO(\bfalpha)=r>0$, that holds one of the conditions from the theorem. Denote $\bfalpha'=\bfalpha_{[1;\ell-r]}$. Let $\boldsymbol{x}\in\Sigma_q^n$ and let $\bfy = L_{\bfalpha}(\bfx)$. It holds that if $\bfx_{[i;\ell]} = \bfalpha$, then $y_{i} = 1$. So, if $\bfx_{[i;k(\ell-r)]}$ consists of a repetition of $\bfalpha'$ $k\geq1$ times, $\bfx_{[k(\ell-r);r]}=\bfalpha_{[1;r]}$, and $\bfx_{[k(\ell-r);\ell]} \neq \bfalpha$, then $\bfy_{[i;k(\ell-r)+r]}$ is of the form $(10^{\ell-r-1})^{k-1}10^{\ell-1}$. In contrarily to the case of a periodic label, $y_{i+k(\ell-r)+r}$ does not have to be zero. As a result, every valid $\bfalpha$-labeling sequence is a binary sequence of the form $(0+(10^{\ell-r-1})^*(10^{\ell-1}))^*$. Denote this set by $S_r$ and $S_r(n) \triangleq S_r\cap\Sigma_2^n$. Next, we show $S_r(n)\subseteq F_n(\bfalpha)$. Let $\bfu\in\Sigma_2^n$ be a binary sequence in $S_r(n)$. Let $\bfv\in\Sigma_q^n$ be a sequence in which if $u_i=1$ then $\bfv_{[i;\ell]}=\bfalpha$. It holds that $L_{\bfalpha}(\bfv)=\bfu$. So, $|F_n(\bfalpha)|=|S_r(n)|$. Denote the constraint that is presented in~\Cref{CyclicOverlapGraph} by $S$.

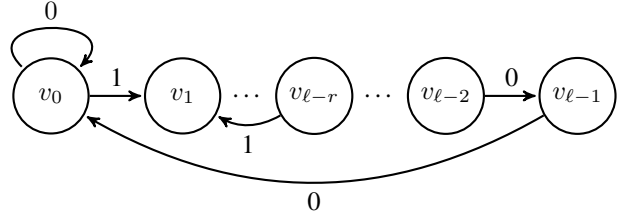
\begin{figure}[h]
\centering
    \begin{tikzpicture}[->,>=stealth',shorten >=1pt,thick,round/.style={draw,circle,thick,minimum size=10mm,thick},]
\SetGraphUnit{3} 
\node[round] at (0,0) (A) [] {$v_0$};
\node[round] (B) at (1.75,0) [] {$v_1$};
\node[round] (D) at (3.5,0) [] {$v_{\ell-r}$};
\node[round] (E) at (5.25,0) [] {$v_{\ell-2}$};
\node[round] (F) at (7,0) [] {$v_{\ell-1}$};

\Loop[dist=1cm,dir=NO,label=$0$,labelstyle=above](A)
\draw[->] (A) to [right] node [above] {1} (B);
\draw[->] (D) to [bend left] node [below] {1} (B);
\node at ($(B)!.5!(D)$) {\ldots};
\node at ($(D)!.5!(E)$) {\ldots};
\draw[->] (E) to [right] node [above] {0} (F);
\draw[->] (F) to [bend left] node [below] {0} (A);
\end{tikzpicture}
\caption{Graph presentation of the constraint in~\Cref{CyclicOverlapTheorem}.}
\label{CyclicOverlapGraph}

\end{figure}

The missing edges of the graph are labeled with $0$. From the structure of the graph, for any sequence $\bfy'$ of length $n-(\ell-1)$, when $n\geq \ell-1$, that is generated by the graph, it holds that ${\bfy\triangleq\bfy'\circ0^{\ell-1}\in S_r(n)\subseteq S(n)}$. Hence, there is an injection from $S(n-(\ell-1))$ to $S_r(n)$ and ${|S(n-(\ell-1))|\leq|S_r(n)|\leq|S(n)|}$. Hence, $\mathsf{cap}(\underline{\bfalpha})= \mathsf{cap}(S)$. The adjacency matrix of this graph is of the following form, when the row in the middle in the illustration below of the matrix $A$ is the $(\ell-r+1)$-th row. 

$$A= \begin{pmatrix}
1 & 1 & 0 & 0 & \ldots &0 & \ldots & 0\\
0 & 0 & 1 & 0 & \ldots &0 & \ldots & 0\\
0 & 0 & 0 & 1 & \ldots &0 & \ldots & 0\\
\vdots & \vdots & \vdots & \vdots & \ddots & \vdots  & \vdots & \vdots \\
0 & 1 & 0 & 0 & \ldots & 1 & \ldots & 0\\
\vdots & \vdots & \vdots & \vdots & \vdots & \vdots & \ddots & \vdots\\
0 & 0 & 0 & 0 & \ldots & 0 & \ldots & 1\\
1 & 0 & 0 & 0 & \ldots & 0 & \ldots & 0\\
\end{pmatrix}.$$

It holds that $A_{i,j}=1$ for $(i,j)$ such that: $i=0, j=0$ or $i=j-1$ or $i=\ell-r,j=1$ or $i=\ell-1,j=0$. Otherwise, $A_{i,j}=0$. It can be shown that the characteristic polynomial of this matrix is: $x^{\ell}-x^{\ell-1}-x^{r}+x^{r-1}-1$. Thus, $\mathsf{cap}(S)=\log_2(\lambda)$ when $\lambda$ is the largest real root of this polynomial.
\end{IEEEproof}

The following corollary shows that for two labels of the same length, that satisfy one of the conditions in~\Cref{CyclicOverlapTheorem}, the larger the overlap is, the larger the labeling capacity is.

\begin{corollary}\label{Ordering overlap}
    Let $\bfalpha_1, \bfalpha_2$ be two non-periodic labels of length $\ell$ with a cyclic overlap $0< \cO(\bfalpha_1)< \cO(\bfalpha_2)$. Moreover, for each label, one of the following conditions holds:
\begin{enumerate}
    \item The cyclic overlap does not contain a cyclic overlap.
    \item The label is an almost periodic label and the almost-periodic suffix of the label does not contain a cyclic overlap.
\end{enumerate}
Then, $\mathsf{cap}(\bfalpha_1)< \mathsf{cap}(\bfalpha_2)$.
\end{corollary}

\begin{IEEEproof}
    Let $\ell>0$, $\bfalpha_1$ be a non periodic label of length $\ell, \cO(\bfalpha_1)=r>0$ and one of the conditions from the corollary holds. From~\Cref{CyclicOverlapTheorem}, it holds that $\mathsf{cap}(\bfalpha_1)=\log_2(\lambda_1)$ when $\lambda_1$ is the largest real root of the polynomial $p(x)=x^{\ell}-x^{\ell-1}-x^{r}+x^{r-1}-1$. Let $\bfalpha_2$ be a non-periodic label of length $\ell$ with cyclic-overlap. Furthermore, assume that $\alpha_2$ satisfies one of the conditions stated in the corollary and that $\cO(\bfalpha_2)=r+k$ for $k\in N$. It holds that $\mathsf{cap}(\bfalpha_2)=\log_2(\lambda_2)$ when $\lambda_2$ is the largest real root of the polynomial $x^{\ell}-x^{\ell-1}-x^{r+k}+x^{r+k-1}-1$. As mentioned earlier, it holds that $\lambda_1,\lambda_2>1$. Using the fact that the function $f(x)=\frac{x^\ell-x^{\ell-1}-1}{x^{r-1}(x-1)}= x^{\ell-r}-\frac{1}{x^{r-1}(x-1)}$ is an increasing function for $x>1$, it can be proven in a similar way to the proof of~\Cref{Ordering periodic} that $\lambda_2>\lambda_1$ and so, $\mathsf{cap}(\bfalpha_2)=\log_2(\lambda_2)>\log_2(\lambda_1)=\mathsf{cap}(\bfalpha_1)$. 
\end{IEEEproof}

Lastly, we consider the case of a non-periodic label with cyclic-overlap, and the period of the cyclic-overlap is $p=1$. For example, the cyclic-overlap of the label $AACAA$ is $AA$ and $\cP(AA)=1$.
First, an example for this case is presented and afterwards the general theorem is shown.

\begin{example}
    Let $\bfalpha=AATAA$, so $\cO(\bfalpha)=2$ and $\cP(AA)=1$. Let $\boldsymbol{x}\in\{A,C,G,T\}^n$ and let $\boldsymbol{y}\in\Sigma_2^n$ be the $\bfalpha$-labeling sequence of $\boldsymbol{x}$, i.e., $\bfy = L_{\bfalpha}(\bfx)$. From the definition of $\bfalpha$-labeling sequences, it holds that if $\bfx_{[i;5]} = AATAA$, then $y_{i} = 1$. So, from the structure of the label, there are four options for the number of zeroes between consecutive ones:
    \begin{enumerate}
        \item If $\bfx_{[i;8]}= AATAATAA$, then $\bfy_{[i;8]}= 10010000$.
        \item If $\bfx_{[i;9]}= AATAAATAA$, then $\bfy_{[i;9]}= 100010000$.
        \item If $\bfx_{[i;10]}= AATAAAATAA$, then $\bfy_{[i;10]}= 1000010000$.
        \item Else, if $\bfx_{[i;5]}= AATAA$, the number of zeroes between $y_i$ and the next one will be larger than $4$.
    \end{enumerate}
    In general, the number of zeroes after a one in $\bfy$ is at least $\ell-\cO(\bfalpha)-1=2$, except for the last one that has to have at least $\ell-1$ zeroes afterwards. In terms of calculating the capacity, we can ignore the last one in the $\bfalpha$-labeling sequences. As a result, this is exactly the case of a non-cyclic label of length $\ell-r=3$.
\end{example}
The last example shows that $$\mathsf{cap}(AATAA)= \mathsf{cap}(AAT)=\mathsf{cap}(ACG).$$

\begin{theorem}\label{Overlap period 1}
    Let $\bfalpha$ be a non-periodic label of length $\ell$ with periodic cyclic-overlap $\cO(\bfalpha)=r>0$ that has period of $1$. Then, $\mathsf{cap}(\bfalpha)= \mathsf{cap}(\bfbeta)$ when $\bfbeta$ is a non-cyclic label of length $\ell-r$.
\end{theorem}

\begin{IEEEproof}
    Let $\bfalpha$ be a non-periodic label of length $\ell$ with periodic cyclic-overlap $\cO(\bfalpha)=r>0$ that has period of $1$. Let $\boldsymbol{x}\in\{A,C,G,T\}^n$ and let $\bfy = L_{\bfalpha}(\bfx)$. It holds that if $\bfx_{[i;\ell]} = \bfalpha$, then $y_{i} = 1$. From the structure of the label, the number of zeroes after a one in $\bfy$ is at least $\ell-r-1$, except for the last one that has to have at least $\ell-1$ zeroes afterwards. In terms of calculating the capacity, we can ignore the last one in the $\bfalpha$-labeling sequences. As a result, this is exactly the case of a non-cyclic label of length $\ell-r$.
\end{IEEEproof}

In order to summarize some of the results from this subsection, the previous theorems are presented in~\Cref{table:1}. In this table, the labels are of length $\ell$, the size of the cyclic-overlap is denoted by $r$ and the size of the period is denoted by $p$. The labeling capacity of each type of label is $\log\lambda$ when $\lambda$ is the largest real root of the suitable polynomial from the table.
\begin{table}[h!]
\centering
\caption{The labeling capacity of a single label.}
\begin{tabular}{|p{2.9cm}|p{1cm}|p{3.7cm}|} 
 \hline
 Label properties & Example & Polynomial \\ [0.5ex] 
 \hline\hline
 non-cyclic & $ACG$ & $x^\ell-x^{\ell-1}-1$ \\ 
 non-cyclic period & $CGCG$ & $x^{\ell+1}-x^\ell-x^{\ell-p+1}+x^{\ell-p}-1$ \\
 non-periodic with non-cyclic overlap & $ATA$ & $x^\ell-x^{\ell-1}-x^r+x^{r-1}-1$ \\
 almost periodic with non-cyclic suffix & $ACACA$ & $x^\ell-x^{\ell-1}-x^r+x^{r-1}-1$ \\
 non-periodic with periodic overlap of period 1 & $AATAA$ &  $x^{\ell-r}-x^{\ell-r-1}-1$\\ [1ex] 
 \hline
\end{tabular}
\label{table:1}
\end{table}

Lastly, we note that there are two other types of labels that are not being discussed in this work, the periodic labels with cyclic-overlap and  the cyclic labels who do not hold the conditions from~\Cref{CyclicOverlapTheorem}.  These cases can be solved in a similar method to the one that was studied in this section and are thus left as an exercise to the reader.

\subsection{Ordering the Labels by Labeling Capacity}
Next, we order the different labels by their corresponding labeling capacity. In order to list all the labels by their labeling capacity, the maximal and minimal labeling capacity values, for labels of length $\ell$, will be shown. 
Moreover, types of labels that achieve the largest labeling capacity and the smallest labeling capacity will be presented.

The following theorem will show that the only type of labels that achieve the smallest labeling capacity are the non-cyclic labels.
\begin{theorem}\label{Smallest capacity}
    Let $\ell> 0$. The only labels of length $\ell$ for which $\mathsf{cap}(\bfalpha)$ is minimized are the non-cyclic labels. That is, $\min_{|\bfalpha|=\ell}\mathsf{cap}(\bfalpha)= \log\lambda$ where $\lambda$ is the largest real root of $x^{\ell}-x^{\ell-1}-1$.
\end{theorem}

\begin{IEEEproof}
    Let $\ell>0$, $\bfalpha_1\neq \bfalpha_2$ be two labels of length $\ell$ for which $\cO(\bfalpha_1)=0$ and $\cO(\bfalpha_2)=r>0$. Denote the first letter of $\bfalpha_2$ by $c$. Let $\widehat{L}_{\bfalpha_1}(\bfx)$ be the complete-$\bfalpha_1$-labeling sequence of a sequence $\bfx$. It holds that every run of ones in $\widehat{L}_{\bfalpha_1}(\bfx)$ is of length $k\cdot\ell$ for $k\geq0$. So, define $\bfx'$ such that for every run of ones in $\widehat{L}_{\bfalpha_1}(\bfx)$ of length $m=k\cdot\ell$ that starts at index $i$, $\bfx'_{[i;m]}=\bfalpha_2\circ\bfalpha_2\circ\cdots\circ\bfalpha_2$, $\bfx'_{i+m}\neq c$ and else, $\bfx'_{i}\neq c$. From the definition of complete-$\bfalpha_2$-labeling sequences, it holds that $\widehat{L}_{\bfalpha_2}(\bfx')= \widehat{L}_{\bfalpha_1}(\bfx)$. Thus, every complete-$\bfalpha_1$-labeling sequence is also a complete-$\bfalpha_2$-labeling sequence and so $\widehat{F}_n(\bfalpha_1)\subseteq \widehat{F}_n(\bfalpha_2)$. For non-cyclic labels, there is a bijection between $F_n(\bfalpha)$ and $\widehat{F}_n(\bfalpha)$, so $|\widehat{F}_n(\bfalpha_1)|= |F_n(\bfalpha_1)|$. Moreover, for $\bfalpha_2$, as for any label, it holds that $|\widehat{F}_n(\bfalpha_2)|\leq |F_n(\bfalpha_2)|$. As a result, $|F_n(\bfalpha_1)|= |\widehat{F}_n(\bfalpha_1)| \leq |\widehat{F}_n(\bfalpha_2)|\leq |F_n(\bfalpha_2)|$. From the definition of labeling capacity, $\mathsf{cap}(\bfalpha_2)\geq \mathsf{cap}(\bfalpha_1)$.
    Now, in order to show that this bound is tight, let us compare between the achievable labeling capacity using non-cyclic labels and cyclic labels of different types. 
    \begin{itemize}
        \item Periodic labels: It is easy to verify that for periodic labels, the smallest labeling capacity is obtained using labels with non-cyclic period. So, the case of non-cyclic period will be discussed. Assume $\bfalpha_2$ is a periodic label, $\cP(\bfalpha_2)=p<\ell$ when the period is non-cyclic. From~\Cref{non-cyclic}, $\mathsf{cap}(\bfalpha_1)=\log_2\lambda_1$ when $\lambda_1>1$ is the largest real root of $q(x)= x^{\ell}-x^{\ell-1}-1$. From~\Cref{periodicTheorem}, $\mathsf{cap}(\bfalpha_2)=\log_2\lambda_2$ when $\lambda_2>1$ is the largest real root of $p(x)= x^{\ell+1}-x^{\ell}-x^{\ell-p+1}+x^{\ell-p}-1 = x(x^\ell-x^{\ell-1}-1)+(1-x^{\ell-p})(x-1)$. Denote $f(x)=(1-x^{\ell-p})(x-1)$, so $p(x)=x\cdot q(x)+f(x)$. Since $\lambda_1>1$, we have that $f(\lambda_1)<0$. Additionally, $q(\lambda_1)=0$ and hence, $p(\lambda_1)< 0$. Moreover, since $\lambda_2>1$, $f(\lambda_2)<0$ and $p(\lambda_2)=0$, thus we have that $q(\lambda_2)>0$. The function $q(x)$ is an increasing function, which implies that $\lambda_2> \lambda_1$, and so $\mathsf{cap}(\bfalpha_1)< \mathsf{cap}(\bfalpha_2)$.
        \item Non-periodic labels with cyclic overlap: It is easy to verify that for non-periodic labels with cyclic overlap, the smallest labeling capacity is obtained when the overlap is non-cyclic. So, this is the case that will be discussed. Assume $\bfalpha_2$ is a non-periodic label with cyclic overlap which is not cyclic. It has been shown in~\Cref{Ordering overlap} that the smallest capacity is obtained when $\cO(\bfalpha_2)=1$. It will be shown in~\Cref{Non-cyclic and overlap 1} that $\mathsf{cap}(\bfalpha_2)= \mathsf{cap}(\bfbeta)$ when $\bfbeta$ is a non-cyclic label of length $\ell-1$ \footnote{Note that this can also be deduced by a small modification to~\Cref{Overlap period 1}.}. From~\Cref{Different length non cyclic}, $\mathsf{cap}(\bfalpha_1)<\mathsf{cap}(\bfbeta)$ and so $\mathsf{cap}(\bfalpha_1)< \mathsf{cap}(\bfalpha_2)$.
    \end{itemize}
\end{IEEEproof}

~\Cref{period 1 and shorter non cyclic} shows a connection between non-cyclic labels and periodic labels with period one. This lemma will be helpful in order to discuss labels that gain the maximal labeling capacity. It shows that, for example, $\mathsf{cap}(ACG)=\mathsf{cap}(AAAAA)$.

\begin{lemma}\label{period 1 and shorter non cyclic}
    Let $\bfalpha_1$ be a non-cyclic label of length $\ell$ and $\bfalpha_2$ be a periodic label with period $\cP(\bfalpha_2)=1$ and length $2\ell-1$. Then, it holds that $\mathsf{cap}(\bfalpha_1)= \mathsf{cap}(\bfalpha_2)$.
\end{lemma}

\begin{IEEEproof}
    Let $\bfalpha_1$ be a non-cyclic label of length $\ell$. From~\Cref{non-cyclic}, $\mathsf{cap}(\bfalpha_1)=\log_2\lambda_1$ when $\lambda_1$ is the largest real root of $x^\ell-x^{\ell-1}-1$. Let $\bfalpha_2$ be a periodic label with period $\cP(\bfalpha_2)=1$ of length $2\ell-1$. From~\Cref{periodicTheorem}, $\mathsf{cap}(\bfalpha_2)=\log_2(\lambda_2)$ when $\lambda_2$ is the largest real root of the polynomial $x^{2\ell}-2x^{2\ell-1}+x^{2\ell-2}-1 = (x^\ell-x^{\ell-1}-1)(x^\ell-x^{\ell-1}+1)$. It has been shown in~\Cref{Different length non cyclic} that $p(x)=x^\ell-x^{\ell-1}-1$ has a real root which is larger than $1$. Moreover, $x^\ell-x^{\ell-1}+1 = x^{\ell-1}(x-1)+1$ has no real roots which are larger than one. So, $\lambda_1=\lambda_2$ and $\mathsf{cap}(\bfalpha_1)= \mathsf{cap}(\bfalpha_2)$.
\end{IEEEproof}

The last lemma and~\Cref{Overlap period 1} show the following corollary, which implies, for example, that $$\mathsf{cap}(AAAAA)=\mathsf{cap}(AACAA)=\mathsf{cap}(ACG).$$
\begin{corollary}\label{periodicNoncyclicAlmostperiodic}
    Let $\bfalpha_1$ be a non-periodic label of length $\ell$, $\bfalpha_2$ be a periodic label of length $2\ell-1$ with period $\cP(\bfalpha_2)=1$ and $\bfalpha_3$ be a non-periodic label of length $2\ell-1$ with periodic cyclic-overlap $\cO(\bfalpha_3)=\ell-1>0$ that has period of 1. Then, $\mathsf{cap}(\bfalpha_1)= \mathsf{cap}(\bfalpha_2)= \mathsf{cap}(\bfalpha_3)$.
\end{corollary}

The following two lemmas are dealing with comparing the labeling capacity of labels with different properties. In~\Cref{Non-cyclic and overlap 1} a connection between non-cyclic labels and labels with cyclic overlap of size one is being shown.~\Cref{cyclic and periodic} shows a connection between periodic labels and non-periodic labels with cyclic overlap.
The next lemma implies that, for example, $\mathsf{cap}(ACG)=\mathsf{cap}(ACGA)$.
 
\begin{lemma}\label{Non-cyclic and overlap 1}
    Let $\bfalpha_1$ be a non cyclic label of length $\ell>1$ and denote the first letter of $\bfalpha_1$ by $c$. Let $\bfalpha_2=\bfalpha_1\circ c$. Then, it holds that $\mathsf{cap}(\bfalpha_1)=\mathsf{cap}(\bfalpha_2)$.
\end{lemma}
\begin{IEEEproof}
    Let $\ell>1$ and let $\bfalpha_1,\bfalpha_2$ be two labels that hold the properties from the lemma.
    Note that $\bfalpha_2$ is a non-periodic label of length $\ell+1$ with a non-cyclic overlap, $\cO(\bfalpha_2)=1$. From ~\Cref{CyclicOverlapTheorem}, it holds that $\mathsf{cap}(\bfalpha_2)=\log_2(\lambda_2)$ when $\lambda_2$ is the largest real root of the polynomial $x^{\ell+1}-x^{\ell}-x= x(x^\ell-x^{\ell-1}-1)$. From ~\Cref{non-cyclic}, it holds that $\mathsf{cap}(\bfalpha_1)=\log_2(\lambda_1)$ when $\lambda_1$ is the largest real root of the polynomial $x^\ell-x^{\ell-1}-1$.
\end{IEEEproof}

\begin{lemma}\label{cyclic and periodic}
    Let $\bfalpha_1$ be a periodic label of length $\ell$ with non-cyclic period $\cP(\bfalpha_2)=p<\ell$ and let $\bfalpha_2$ be a label of length $\ell+1$ with a cyclic-overlap $\cO(\bfalpha_1)=r=\ell-p+1$ that holds the conditions from~\Cref{CyclicOverlapTheorem}. It holds that $\mathsf{cap}(\bfalpha_1)= \mathsf{cap}(\bfalpha_2)$.
\end{lemma}

\begin{IEEEproof}
    From~\Cref{periodicTheorem}, $\mathsf{cap}(\bfalpha_1)=\log_2(\lambda_1)$ when $\lambda_1$ is the largest real root of $p(x)= x^{\ell+1}-x^{\ell}-x^{\ell-p+1}+x^{\ell-p}-1$. Additionally, from~\Cref{CyclicOverlapTheorem}, $\mathsf{cap}(\bfalpha_2)=\log_2(\lambda_2)$ when $\lambda_2$ is the largest real root of $p(x)$.
\end{IEEEproof}

In the following theorem, labels that achieve the largest capacity are discussed. For $\ell\leq 5$, we show one type of labels that achieve this maximum for all lengths of labels. For the case in which the labels' length is odd, we present an additional type of label that also achieve the maximum capacity. For example, it is shown that, for $\ell=5$, the labels with the largest labeling capacity are of the following forms: $AAAAA,AACAA$. For $\ell=4$, the labels are of the following form: $AAAA$.
\begin{theorem}\label{Largest capacity}
    Let $\ell\leq 5$ and let $\sigma, \tau\in \Sigma_q$ be two distinct symbols. For odd $\ell$, two types of labels of length $\ell$ for which $\mathsf{cap}(\bfalpha)$ is maximized are $\bfalpha=\sigma^\ell$, where $\sigma^\ell$ is the label that consists of $\ell$ repetitions of the letter $\sigma$ and $\sigma^{\frac{\ell-1}{2}}\tau\sigma^{\frac{\ell-1}{2}}$. For even $\ell$, one type of labels of length $\ell$ for which $\mathsf{cap}(\bfalpha)$ is maximized is $\bfalpha=\sigma^\ell$. So, for any $\ell\le 5$, $\max_{|\bfalpha|=\ell}\mathsf{cap}(\bfalpha)= \mathsf{cap}(\sigma^\ell)= \log\lambda$ where $\lambda$ is the largest real root of $x^{\ell+1}-2x^{\ell}+x^{\ell-1}-1$.
\end{theorem}
\begin{IEEEproof}
    Let $\ell\leq5$ and let $\sigma,\tau\in \Sigma_q$ be two distinct symbols. Assume w.l.o.g. $\sigma=A, \tau=C$. From~\Cref{periodicNoncyclicAlmostperiodic} it is enough to prove the theorem for $A^\ell$ since for odd $\ell$, $\mathsf{cap}(A^\ell)= \mathsf{cap}(A^{\frac{\ell-1}{2}}CA^{\frac{\ell-1}{2}})$.
    Let $\bfalpha_1=A^\ell$ and $\bfalpha_2\neq \bfalpha_1$ be two labels of length $\ell$. If $\bfalpha_2=\sigma_0^\ell$ or $\bfalpha_2=\sigma_0^{\frac{\ell-1}{2}}\tau_0\sigma_0^{\frac{\ell-1}{2}}$ for two distinct symbols $\sigma_0, \tau_0\in \Sigma_q$, it can be verified that $\mathsf{cap}(A^\ell)= \mathsf{cap}(\bfalpha_2)$. From~\Cref{Smallest capacity}, if $\bfalpha_2$ is a non-cyclic label, $\mathsf{cap}(\bfalpha_2)< \mathsf{cap}(\bfalpha_1)$. Moreover, since $\bfalpha_1$ is periodic with non-cyclic period, from~\Cref{Ordering periodic}, if $\bfalpha_2$ is also periodic with non-cyclic period, then $\mathsf{cap}(\bfalpha_2)< \mathsf{cap}(\bfalpha_1)$. There is no periodic label of length $\ell\leq5$ with cyclic period. Thus, the only case that remains to be studied is the one where $\bfalpha_2$ is a non-periodic label with cyclic-overlap.
    Assume $\bfalpha_2$ is a non-periodic label with a cyclic-overlap $\cO(\bfalpha_2)=r>0$.
    There are three different cases:
    \begin{enumerate}
        \item $r=\frac{\ell}{2}$: This case is not possible, since the label is not periodic.
        \item $r<\frac{\ell}{2}$: Let $\bfy= L_{\bfalpha_2}(\bfx)$ be the $\bfalpha_2$-labeling sequence of a sequence $\bfx$. From the structure of $\bfalpha_2$, there must be at least $\ell-r-1$ zeroes between two consecutive ones in $\bfy$. If this is the only constraint, the result is that $\bfy$ satisfies the $(\ell-r-1,\infty)$-RLL constraint. Let $\bfalpha_3$ be a non-cyclic label of length $\ell-r$. From~\Cref{non-cyclic}, $\mathsf{cap}(\bfalpha_3)$ is equal to the capacity of the $(\ell-r-1,\infty)$-RLL constraint. 
        So, $\mathsf{cap}(\bfalpha_3)\geq\mathsf{cap}(\bfalpha_2)$. From~\Cref{Different length non cyclic}, the greater the value of $r$ is, the greater the labeling capacity of $\bfalpha_3$ is. So, let $\bfalpha_4$ be a non-cyclic label of length $\ell-r^*$ when $r^*=\frac{\ell-1}{2}$ if $\ell$ is odd and $r^*=\frac{\ell-2}{2}$ if $\ell$ is even. It holds that $\mathsf{cap}(\bfalpha_4)\geq\mathsf{cap}(\bfalpha_3)\geq\mathsf{cap}(\bfalpha_2)$. From~\Cref{periodicNoncyclicAlmostperiodic}, if $\ell$ is odd, then $\mathsf{cap}(\bfalpha_1)=\mathsf{cap}(\bfalpha_4)$, so $\mathsf{cap}(\bfalpha_1)\geq\mathsf{cap}(\bfalpha_2)$. If $\ell$ is even, from~\Cref{periodicNoncyclicAlmostperiodic}, $\mathsf{cap}(A^{\ell+1})=\mathsf{cap}(\bfalpha_4)$. Lastly, from~\Cref{Different length same period}, $\mathsf{cap}(A^{\ell+1})<\mathsf{cap}(A^{\ell})$ and so, $\mathsf{cap}(A^{\ell})>\mathsf{cap}(\bfalpha_2)$.
        \item $r>\frac{\ell}{2}$: This case can happen only if $\bfalpha_2$ is almost-periodic as defined in~\Cref{almostPeriodicDef} and $\ell>3$. Additionally, if $\ell=4$, it must be that $r=3$, which is possible only if $\bfalpha_2$ is periodic, a contradiction. Hence $\ell= 5$ and the almost-periodic suffix of $\bfalpha_2$ does not contain a cyclic overlap. 
        From~\Cref{Ordering overlap}, the greater the value of $r$ is, the greater the labeling capacity of $\bfalpha_2$ is. Note that $r<\ell-1$ since otherwise the label is periodic. So, assume $r=\ell-2$.
        From~\Cref{CyclicOverlapTheorem}, $\mathsf{cap}(\bfalpha_2)=\log_2(\lambda_2)$ when $\lambda_2$ is the largest real root of the polynomial $p(x)= x^{\ell}-x^{\ell-1}-x^{r}+x^{r-1}-1$. The largest root of $p(x)$ is also the largest root of 
        \begin{align*}
            x\cdot p(x)= x^{\ell+1}-x^{\ell}-x^{r+1}+x^{r}-x= (x^{\ell+1}-2x^{\ell}+x^{\ell-1}-1)+(x^{\ell}-x^{\ell-1}-x^{r+1}+x^{r}-x+1)= \\(x^{\ell+1}-2x^{\ell}+x^{\ell-1}-1)+(x-1)(x^{\ell-1}-x^r-1)=(x^{\ell+1}-2x^{\ell}+x^{\ell-1}-1)+(x-1)(x^{\ell-1}-x^{\ell-2}-1).
        \end{align*}
        Denote $a(x)= x^{\ell+1}-2x^{\ell}+x^{\ell-1}-1$ and $b(x)= x^{\ell-1}-x^{\ell-2}-1$.
        From~\Cref{periodicTheorem}, $\mathsf{cap}(A^\ell)=\log_2(\lambda_1)$ when $\lambda_1$ is the largest real root of $a(x)$. From~\Cref{periodicNoncyclicAlmostperiodic}, $\mathsf{cap}(A^\ell)$ is equal to the labeling capacity of a non-cyclic label of length $\frac{\ell+1}{2}$ if $\ell$ is odd and is greater than the labeling capacity of a non-cyclic label of length $\frac{\ell+2}{2}$ if $\ell$ is even. 
        From~\Cref{non-cyclic}, the labeling capacity of a non-cyclic label of length $\ell-1$ is $\log_2(\lambda)$ when $\lambda$ is the largest real root of $b(x)$. 
        For $\ell>3$ it holds that $\frac{\ell+1}{2}<\ell-1$ if $\ell$ is odd and $\frac{\ell+2}{2} \leq \ell-1$ if $\ell$ is even. So, from~\Cref{Different length non cyclic}, $\lambda_1 \geq \lambda$. 
        Thus, from the structure of $x\cdot p(x)$, since $a(x),b(x)$ are increasing functions for $x>1$ it holds that $\lambda< \lambda_2< \lambda_1$ and $\mathsf{cap}(A^{\ell})>\mathsf{cap}(\bfalpha_2)$.
    \end{enumerate}
\end{IEEEproof}

\subsection{Ordering all the Labels of Length $\ell\leq 5$}
Lastly, the order of all labels of length $\ell\leq5$, according to their labeling capacity values is shown.
\begin{enumerate}
    \item It is easy to see that the largest capacity is obtained for labels of length $1$. 
    \item For labels of length $2$ there are two cases:
    \begin{enumerate}
        \item The periodic label with period $1$, for example $AA$: From ~\Cref{Largest capacity}, this label achieves the largest capacity for labels of length $2$.
        \item The non-cyclic label, for example $AC$: From ~\Cref{Smallest capacity}, this label achieves the smallest capacity for labels of length~$2$. 
    \end{enumerate}
    So, it holds that $$\mathsf{cap}(AC) < \mathsf{cap}(AA) < \mathsf{cap}(A).$$
    \item For labels of length $3$ there are three cases:
        \begin{enumerate}
        \item The periodic label with period $1$, for example $AAA$: From~\Cref{Largest capacity}, this label achieves the largest capacity for labels of length $3$. From~\Cref{Different length same period}, $\mathsf{cap}(AAA) < \mathsf{cap}(AA)$. Moreover, from~\Cref{period 1 and shorter non cyclic}, $\mathsf{cap}(AAA)= \mathsf{cap}(AC)$.
        \item The non-periodic label with non-cyclic overlap of size $1$, for example $ACA$: From~\Cref{Non-cyclic and overlap 1}, the labeling capacity of this label is the same as the labeling capacity of the label $AC$.
        \item The non cyclic label, for example $ACG$: From ~\Cref{Smallest capacity}, this label achieves the smallest capacity for labels of length~$3$. 
    \end{enumerate}
    Thus, 
    \begin{align*}
    & \mathsf{cap}(ACG) <  \mathsf{cap}(ACA) =\mathsf{cap}(AC) \\ 
    & = \mathsf{cap}(AAA) < \mathsf{cap}(AA) < \mathsf{cap}(A).
    \end{align*}
    \item For labels of length $4$ there are four cases:
        \begin{enumerate}
        \item The periodic label with period $1$, for example $AAAA$ that achieves the largest capacity for labels of length $4$. From~\Cref{Different length same period}, $\mathsf{cap}(AAAA) < \mathsf{cap}(AAA)$.
        \item The periodic label with period $2$, for example $ACAC$. 
        \item The non-periodic label with non-cyclic overlap of size $1$, for example $ACGA$: From~\Cref{Non-cyclic and overlap 1}, the labeling capacity of this label is the same as the labeling capacity for the label $ACG$. Moreover, from the proof of~\Cref{Largest capacity}, $\mathsf{cap}(ACGA) < \mathsf{cap}(AAAA)$.
        \item The non cyclic label, for example $ACGT$ that achieves the smallest capacity for labels of length $4$. 
    \end{enumerate}
    So, 
    \begin{align*}
      & \mathsf{cap}(ACGT)< \mathsf{cap}(ACG) = \mathsf{cap}(ACGA) \\
      & \leq \mathsf{cap}(AAAA) < \mathsf{cap}(ACA) = \mathsf{cap}(AC) \\ 
      & = \mathsf{cap}(AAA) < \mathsf{cap}(AA) < \mathsf{cap}(A).
    \end{align*}
%    $$ $$$$ $$$$$$
    Moreover, from~\Cref{Ordering periodic}, $\mathsf{cap}(ACAC)$ $< \mathsf{cap}(AAAA)$ but in order to insert the label $ACAC$ to the ordering above, the labels of length $5$ will be discussed first.
        \item For labels of length $5$ there are six cases:
        \begin{enumerate}
        \item The periodic label with period $1$, for example $AAAAA$ that achieves the largest capacity for labels of length $5$. From~\Cref{period 1 and shorter non cyclic}, $\mathsf{cap}(AAAAA) = \mathsf{cap}(ACG)$.
        \item The non-periodic label with periodic overlap with periodicity $1$ of size $2$, for example $AACAA$: From~\Cref{Overlap period 1}, $\mathsf{cap}(AACAA) = \mathsf{cap}(ACG)$.
        \item The non-periodic label with non-cyclic overlap of size $1$, for example $ACGTA$: From ~\Cref{Non-cyclic and overlap 1}, the labeling capacity of this label is the same as the labeling capacity for the label $ACGT$.
        \item The non-periodic label with non-cyclic overlap of size $2$, for example $ACGAC$.
        \item The non-periodic label with cyclic-overlap of size $3$, for example $ACACA$: From~\Cref{cyclic and periodic}, $\mathsf{cap}(ACACA) = \mathsf{cap}(ACAC)$.
        From~\Cref{Ordering overlap}, it holds that $\mathsf{cap}(ACGTA) < \mathsf{cap}(ACGAC) < \mathsf{cap}(ACACA)$. It can be shown by using~\Cref{periodicTheorem} and~\Cref{CyclicOverlapTheorem} that $\mathsf{cap}(ACACA) < \mathsf{cap}(AAAAA)$.
        \item The non cyclic label, for example $ACGAT$ that achieves the smallest capacity for labels of length $5$. 
    \end{enumerate}
    \end{enumerate}
    
\begin{figure}[h]
\centering
\begin{tikzpicture}[
    round/.style={draw,thick,text width=1.3cm, align=center},
    ]
    \node[round] at (3,9.5) (1) [] {$A$};
    \node[round] (2) at (5,9) [] {$AA$};
    \node[round] (3) at (7,8) [] {$AAA$ $ACA$};
    \node[round] (6) at (5,8) [] {$AC$};
    \node[round] (4) at (9,7) [] {$AAAA$};
    \node[round] (5) at (11,6) [] {$AAAAA$ $AACAA$};
    \node[round] (7) at (9,6) [] {$ACGA$};
    \node[round] (8) at (7,6) [] {$ACG$};
    \node[round] (9) at (11,5) [] {$ACACA$};
    \node[round] (10) at (9,5) [] {$ACAC$};
    \node[round] (11) at (11,4) [] {$ACGAC$};
    \node[round] (12) at (11,3) [] {$ACGTA$};
    \node[round] (13) at (9,3) [] {$ACGT$};
    \node[round] (14) at (11,2) [] {$ACGAT$};
    \draw[->] (2) to [left] node [above] {} (6);
    \draw[->] (4) to [left] node [above] {} (7);
    \draw[->] (3) to [left] node [above] {} (8);
    \draw[->] (5) to [left] node [above] {} (9);
    \draw[->] (7) to [left] node [above] {} (10);
    \draw[->] (9) to [left] node [above] {} (11);
    \draw[->] (10) to [left] node [above] {} (13);
    \draw[->] (11) to [left] node [above] {} (12);
    \draw[->] (12) to [left] node [above] {} (14);
    \end{tikzpicture}
  \label{notPathUnique}
\caption{The order between different types of labels of length $\ell \leq 5$, according to their labeling capacity values.}
\label{fig:order}
\end{figure}
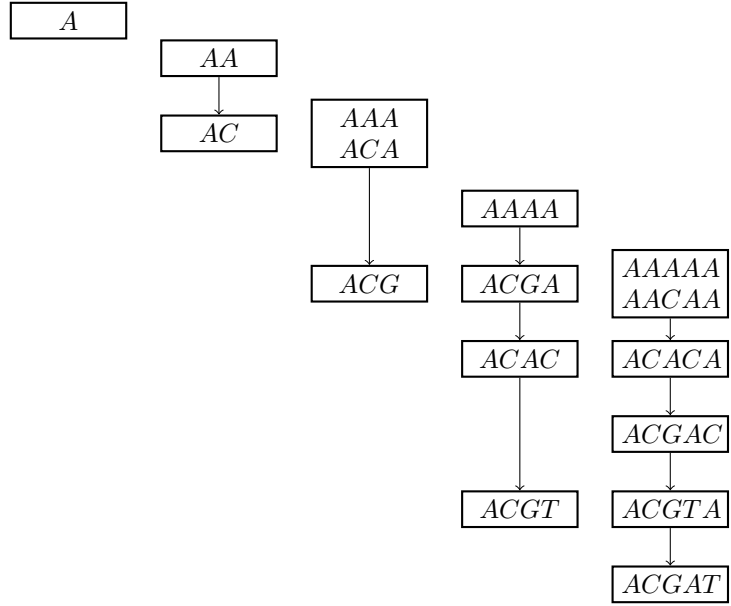

    Finally, these relations are depicted in~\Cref{fig:order}. Each node corresponds to a type of labels. For example, the label $ACA$ corresponds to the non-periodic labels of length three with non-cyclic overlap of size one, as $ATA,CGC,TCT$. The labels in each row have the same labeling capacity. The lower the label is, the smaller the labeling capacity is. All of the inequalities between the labeling capacity values of the different rows are strict. The arrows present the connection between the labels of the same length.
    
\section{The labeling Capacity of Multiple Labels}
\label{sec:mul}

In this section, the labeling capacity in case multiple labels are being used is studied. The next theorem solves the case of $k$ non-overlapping non-cyclic labels. First, an example for this case is presented.

\begin{example}\label{mulLabelsExample}
Let $\bfalpha_1=AC, \bfalpha_2=GT, \bfalpha_3=AGCT$ be non-overlapping non-cyclic labels of lengths $\ell_1=2,\ell_2=2,\ell_3=4$ respectively. Denote $\underline{\bfalpha}=(\bfalpha_1, \bfalpha_2, \bfalpha_3)$. Denote the constraint that is presented in~\Cref{mulLabelsGraph} by $S$.

\begin{figure}[h]
\centering
    \begin{tikzpicture}[->,>=stealth',shorten >=1pt,thick,round/.style={draw,circle,thick,minimum size=10mm,thick},]
\SetGraphUnit{3} 
\node[round] at (0,0) (A) [] {$v_0$};
\node[round] (B) at (2.2,0) [] {$v_1$};
\node[round] (C) at (4.4,0) [] {$v_2$};
\node[round] (D) at (6.6,0) [] {$v_3$};

\Loop[dist=1cm,dir=NO,label=$0$,labelstyle=above](A)
\draw[->] (A) to [right] node [above] {1,2} (B);
\draw[->] (B) to [bend left] node [below] {0} (A);
\draw[->] (C) to [bend left] node [below] {0} (B);
\draw[->] (D) to [bend left] node [below] {0} (C);
\Loop[dist=1cm,dir=NO,label=$0$,labelstyle=above](A)
\draw[->] (A) to [bend left] node [above] {3} (D);
\end{tikzpicture}
\caption{Graph presentation of the constraint in~\Cref{mulLabelsExample}.}
\label{mulLabelsGraph}
\end{figure}
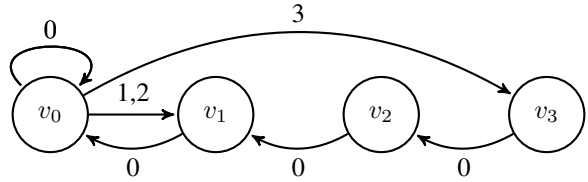

From the structure of the graph, for any sequence $\bfy'$ of length $n-3$, when $n\geq 3$, that is generated by the graph, it holds that ${\bfy\triangleq\bfy'\circ000\in F_n(\underline{\bfalpha})\subseteq S(n)}$. So, it holds that $|S(n-3)|\leq|F_n(\underline{\bfalpha})|\leq|S(n)|$. As a result, $\mathsf{cap}(\underline{\bfalpha})=\mathsf{cap}(S)$.
    
The adjacency matrix of this graph is
$$\begin{pmatrix}
1 & 2 & 0 & 1\\
1 & 0 & 0 & 0\\
0 & 1 & 0 & 0\\
0 & 0 & 1 & 0\\
\end{pmatrix},$$
and its characteristic polynomial is $x^4-x^3-2x^2-1$. Thus, the capacity of $\cS$ is $\log_2(\lambda)$ when $\lambda \approx 2.075$ is the largest real root of this polynomial.
\end{example}

\begin{theorem}\label{multipleNonOverlaping}
Let $\underline{\bfalpha}= (\bfalpha_1,\ldots,\bfalpha_k)$ be $k$ non-overlapping non-cyclic labels of lengths $\ell_1\leq\cdots\leq\ell_k$, respectively. Denote the number of labels of length $j$ by $m_j$. It holds that $\mathsf{cap}(\underline{\bfalpha})=\log_2(\lambda)$ where $\lambda$ is the largest real root of the polynomial $x^{\ell_k}-x^{\ell_k-1}-\sum_{i=1}^{\ell_{k}}m_ix^{\ell_k-i}$.
\end{theorem}

\begin{IEEEproof}
The valid $\underline{\bfalpha}$-labeling sequences are the sequences over $\Sigma_{k+1}$ in which after each $i\in[k]$ there are at least $\ell_i-1$ zeroes.
The valid $\underline{\bfalpha}$-labeling sequences could be presented in a graph $G$, for which the adjacency matrix $A_{\ell_k\times\ell_k}$ will be as follows. For $(i,j)$ such that $i=j+1$, $A_{i,j}=1$. Moreover, $A_{0,0}= 1+ m_1$ and for $j>0$, $A_{0,j}=m_{j+1}$. Otherwise, $A_{i,j}=0$.
It can be shown that the characteristic polynomial of this matrix is $x^{\ell_k}-(1+m_1)x^{\ell_k-1}-\sum_{i=2}^{\ell_k}m_ix^{\ell_k-i}$.
So, it holds that $\mathsf{cap}(\underline{\bfalpha})=\mathsf{cap}(S(G))=\log_2(\lambda)$, where $\lambda$ is the largest real root of this polynomial.
\end{IEEEproof}

Note that in case the $k$ non-overlapping non-cyclic labels are of the same length $\ell$, $\mathsf{cap}(\underline{\bfalpha})=\log_2\lambda$ when $\lambda$ is the largest real root of the polynomial $x^{\ell}-x^{\ell-1}-k$. It has been proven in~\cite{McLaughlin95} that this is the capacity of the $(k+1,\ell-1,\infty)$-RLL constraint.

The next case to be discussed is the one of using overlapping labels. The special case in which the labeling capacity of two non-cyclic labels $\bfalpha_1,\bfalpha_2$, when $\cO(\bfalpha_1,\bfalpha_2)>0, \cO(\bfalpha_2,\bfalpha_1)=0$ will be derived. Additional cases can be studied similarly to the following example and theorem.

\begin{example}
\label{twoLabelsExample}
    Let $\underline{\bfalpha}=(\bfalpha_1,\bfalpha_2)$ when $\bfalpha_1= ACGT, \bfalpha_2= GTT$, are two non-cyclic labels of lengths $\ell_1=4, \ell_2=3$. It holds that $\cO(\bfalpha_1,\bfalpha_2)=t=2, \ \cO(\bfalpha_2,\bfalpha_1)=0$.
    Let $\boldsymbol{x}\in\Sigma_4^n$ and let $\boldsymbol{y}\in\Sigma_3^n$ be the $\underline{\bfalpha}$-labeling sequence of $\boldsymbol{x}$, i.e., $\bfy = L_{\underline{\bfalpha}}(\bfx)$. From the definition, it holds that:
    \begin{itemize}
        \item If $\bfx_{[i;4]} = ACGT$ and $x_{i+4} \neq T$, then $\bfy_{[i;4]} = (1,0,0,0)$. In the general case, if $\bfx_{[i;\ell_1]} = \bfalpha_1$ and $\bfx_{[i+\ell_1;\ell_2-t]} \neq \bfalpha_{2[1+t;\ell_2-t]}$, then $\bfy_{[i;\ell_1]} = (1,0,\ldots,0)$.
        \item If $\bfx_{[i;4]} = ACGT$ and $x_{i+4} = T$, then $\bfy_{[i;5]} = (1,0,2,0,0)$.
        In general, if $\bfx_{[i;\ell_1]} = \bfalpha_1$ and $\bfx_{[i+\ell_1;\ell_2-t]} = \bfalpha_{2[1+t;\ell_2-t]}$, then $\bfy_{[i;\ell_1-t]} = (1,0,\ldots,0)$, $\bfy_{[i+\ell_1-t;\ell_2]} = (2,0,\ldots,0)$.
        \item If $\bfx_{[i;3]} = GTT$ then $\bfy_{[i;3]} = (2,0,0)$. In general, if $\bfx_{[i;\ell_2]} = \bfalpha_2$ then $\bfy_{[i;\ell_2]} = (2,0,\ldots,0)$.
        \item Else, $y_i=0$.
    \end{itemize}
    The correctness is due to the fact that the sequences are non-cyclic.
    So, every valid $\underline{\bfalpha}$-labeling sequence is a ternary sequence in which (1) each one is followed by 3 zeroes or a zero and a two, and (2) each two is followed by at least 2 zeroes.
    
    Denote the set of ternary sequences that hold these two conditions by $S_c$, and $S_c(n)\triangleq S_c\cap\Sigma_3^n$.
    Next, we prove that $S_{c}(n)\subseteq F_n(\underline{\bfalpha})$. Let $\bfu\in S_{c}(n)$ and let $\bfv\in\Sigma_4^n$ be a sequence in which if $\bfu_i=1$, then $\bfv_{[i;4]}= ACGT$ and if $\bfu_i=2$, then $\bfv_{[i;3]}= GTT$. It holds that $L_{\underline{\bfalpha}}(\bfv)=\bfu$.
    So, the valid $\underline{\bfalpha}$-labeling sequences of length $n$ are the sequences in $S_{c}(n)$, which means that $F_n(\underline{\bfalpha})=S_{c}(n)$. Denote the constraint that is presented in~\Cref{twoLabelsExampleGraph} by $S$.

\begin{figure}[h]
\centering
    \begin{tikzpicture}[->,>=stealth',shorten >=1pt,thick,round/.style={draw,circle,thick,minimum size=10mm,thick},]

\node[round] at (0,0) (1) [] {$v_0$};
\node[round] (2) at (2,0) [] {$v_1$};
\node[round] (3) at (4,0) [] {$v_2$};
\node[round] (4) at (6,0) [] {$v_3$};
\node[round] (5) at (2,-2) [] {$v_4$};
\node[round] (6) at (4,-2) [] {$v_5$};

%Lines
\Loop[dist=1cm,dir=NO,label=$0$,labelstyle=above](1)

\draw[->] (1) to [right] node [above] {1} (2);
\draw[->] (2) to [right] node [above] {0} (3);
\draw[->] (3) to [right] node [above] {0} (4);
\draw[->] (4) to [bend right] node [above] {0} (1);

\draw[->] (1) to [bend right] node [below] {2} (5);
\draw[->] (5) to [right] node [below] {0} (6);
\draw[->] (6) to [right] node [above] {0} (1);
\draw[->] (3) to [right] node [above] {2} (5);
\end{tikzpicture}  
\caption{Graph presentation of the constraint in~\Cref{twoLabelsExample}.}
\label{twoLabelsExampleGraph}
\end{figure}
From the structure of the graph, for any sequence $\bfy'$ of length $n-3$, when $n\geq 3$, that is generated by the graph, it holds that ${\bfy\triangleq\bfy'\circ000\in F_n(\underline{\bfalpha})\subseteq S(n)}$. So, $|S(n-3)|\leq|F_n(\underline{\bfalpha})|\leq|S(n)|$. As a result, $\mathsf{cap}(\underline{\bfalpha})= \mathsf{cap}(S)$.
The adjacency matrix of this graph is 

\begin{small}
$$\begin{pmatrix}
1 & 1 & 0 & 0 & 1 & 0\\
0 & 0 & 1 & 0 & 0 & 0\\
0 & 0 & 0 & 1 & 1 & 0\\
1 & 0 & 0 & 0 & 0 & 0\\
0 & 0 & 0 & 0 & 0 & 1\\
1 & 0 & 0 & 0 & 0 & 0\\
\end{pmatrix}$$
\end{small}

\noindent which has the characteristic polynomial $x^6-x^5-x^3-x^2-x$. Thus, the capacity of $S$ is $\log_2(\lambda)$ when $\lambda\approx 1.685$ is
the largest real root of this polynomial.
\end{example}
Next this case is proved formally.
\begin{restatable}{theorem}{twoLabels}
\label{twoLabelsTheorem}
    Let $\bfalpha_1,\bfalpha_2$ be two non-cyclic labels of lengths $\ell_1,\ell_2$ respectively while $\cO(\bfalpha_1,\bfalpha_2)=t>0, \cO(\bfalpha_2,\bfalpha_1)=0$. It holds that $\mathsf{cap}(\bfalpha_1,\bfalpha_2)=\log_2(\lambda)$, where $\lambda$ is the largest real root of $x^{\ell_1+\ell_2-1}-x^{\ell_1+\ell_2-2}-x^{\ell_1-1}-x^{\ell_2-1}-x^{t-1}$.    
\end{restatable}

\begin{IEEEproof}
Let $\underline{\bfalpha} = (\bfalpha_1,\bfalpha_2)$ be two overlapping non-cyclic labels of lengths $\ell_1,\ell_2$ respectively, such that $\cO(\bfalpha_1,\bfalpha_2)=t>0$ and $\cO(\bfalpha_2,\bfalpha_1)=0$. 
Let $\boldsymbol{x}\in\Sigma_q^n$ and let $\bfy = L_{\underline{\bfalpha}}(\bfx)$. The valid $\underline{\bfalpha}$-labeling sequences are the ternary sequences in which the following conditions hold:
\begin{itemize}
    \item Each one is followed by at least $\ell_1-1$ zeroes or $\ell_1-t-1$ zeroes and a two.
    \item Each two is followed by at least $\ell_2-1$ zeroes.
\end{itemize}
The graph from~\Cref{twoLabelsExample} can be generalized to the following graph. Denote the constraint that is presented in~\Cref{twoLabelsGraph} by $S$.

\begin{figure}[h]
\centering
    \begin{tikzpicture}[->,>=stealth',shorten >=1pt,thick,roundnode/.style={draw,circle,thick,minimum size=13mm,thick},]

%Nodes
\node[roundnode] (1) {$v_0$};
\node[roundnode] (2) [right=of 1] {$v_1$};
\node[roundnode] (3) [right=of 2] {$v_{\ell_1-t}$};
%\node[roundnode] (4) [right=of 3] {};
\node[roundnode] (5) [right=of 3] {$v_{\ell_1-1}$};
\node[roundnode] (6) [below=of 2] {$v_{\ell_1}$};
%\node[roundnode] (7) [right=of 6] {};
\node[roundnode] (8) [right=of 6] {$v_{\ell_1+\ell_2-2}$};

%Lines
\Loop[dist=1cm,dir=NO,label=$0$,labelstyle=above](1)

\draw[->] (1) to [right] node [above] {1} (2);
\node at ($(2)!.5!(3)$) {\ldots};
%\draw[->] (3) to [right] node [above] {0} (4);
%\draw[->] (4) to [right] node [above] {0} (5);
\draw[->] (5) to [bend right] node [above] {0} (1);

\draw[->] (1) to [bend right] node [below] {2} (6);
%\draw[->] (6) to [right] node [below] {0} (7);
\node at ($(3)!.5!(5)$) {\ldots};
\node at ($(6)!.5!(8)$) {\ldots};
\draw[->] (8) to [right] node [above] {0} (1);

\draw[->] (3) to [right] node [above] {2} (6);
\end{tikzpicture}  
\caption{Graph presentation of the constraint in~\Cref{twoLabelsTheorem}.}
\label{twoLabelsGraph}
\end{figure}

The missing edges are labeled by a $0$. From the structure of the graph, for any sequence $\bfy'$ of length $n-(\max{(\ell_1,\ell_2)}-1)$, when $n\geq \max{(\ell_1,\ell_2)}-1$, that is generated by the graph, it holds that ${\bfy\triangleq\bfy'\circ0^{\max{(\ell_1,\ell_2)}-1}\in F_n(\underline{\bfalpha})\subseteq S(n)}$. So, $|S(n-(\max{(\ell_1,\ell_2)}-1))|\leq|F_n(\underline{\bfalpha})|\leq|S(n)|$.
As a result, $\mathsf{cap}(\underline{\bfalpha})=\mathsf{cap}(S)$.

If there is no label of length one, the adjacency matrix $A$ of this graph satisfies that $A_{i,j}=1$ for $(i,j)$ such that: $i=0, j=0,1,\ell_1$ or $i=j-1$ if $i\neq\ell_1-1$ or $i=\ell_1-1,\ell_1+\ell_2-2 ,j=0$ or $i=\ell_1-t, j=\ell_1$. Otherwise, $A_{i,j}=0$. If one of the labels is of length one, the first row of the matrix would be different, such that $A_{0,0}=2$, $A_{0,1}=1$ and $A_{0,j}=0$ otherwise. 

It can be shown that the characteristic polynomial of this matrix is: $x^{\ell_1+\ell_2-1}-x^{\ell_1+\ell_2-2}-x^{\ell_1-1}-x^{\ell_2-1}-x^{t-1}$. Thus, the capacity of $S$ is $\log_2(\lambda)$ when $\lambda$ is the largest real root of this polynomial.
\end{IEEEproof}

The last case to be discussed in this section is the case of using two labels of the same length with period of size one. This case will be useful in order to find the labeling capacity that can be gained using two labels of length two in the last section of this paper.

\begin{theorem}\label{multiplePeriodic}
    Let $\underline{\bfalpha}= (\bfalpha_1,\bfalpha_2)$ be two periodic labels of length $\ell>1$ with $\cP(\bfalpha_1)=\cP(\bfalpha_2)=1$. It holds that $\mathsf{cap}(\underline{\bfalpha})=\log_2(\lambda)$ where $\lambda$ is the largest real root of the polynomial $x^{\ell-1}(x-1)(x^{\ell+1}- 2x^\ell +x^{\ell-1}-2)- (x+1)$.
\end{theorem}
\begin{IEEEproof}
    Let $\underline{\bfalpha} = (\bfalpha_1,\bfalpha_2)$ be two different labels of length $\ell$ with periodicity one. Let $\boldsymbol{x}\in\Sigma_q^n$ and let $\bfy = L_{\underline{\bfalpha}}(\bfx)$. The valid $\underline{\bfalpha}$-labeling sequences are the ternary sequences in which each non-zero symbol repeats itself several times (at least once) and is then followed by one of the following two options:
    \begin{enumerate}
        \item $\ell$ zeroes.
        \item $\ell-1$ zeroes and then the other non-zero symbol.
    \end{enumerate}
    The last non-zero symbol in the labeling sequence can be followed by only $\ell-1$ zeroes.
    Denote the set of ternary sequences that hold this condition by $S_{p=1,\ell}$, and $S_{p=1,\ell}(n)\triangleq S_{p=1,\ell}\cap\Sigma_3^n$.
    Next, it will be proven that $S_{p=1,\ell}(n)\subseteq F_n(\underline{\bfalpha})$. Let $\bfu\in S_{p=1,\ell}(n)$ and let $\bfv\in\Sigma_3^n$ be a sequence in which if $\bfu_i=j$, then $\bfv_{[i;\ell]}=\bfalpha_j$. It holds that $L_{\underline{\bfalpha}}(\bfv)=\bfu$.
    So, the valid $\underline{\bfalpha}$-labeling sequences of length $n$ are the sequences in $S_{p=1,\ell}(n)$, which means that $F_n(\underline{\bfalpha})=S_{p=1,\ell}(n)$. Denote the constraint that is presented in~\Cref{multiplePeriodicGraph} by $S$.
\begin{figure}[h]
\centering
    \begin{tikzpicture}[->,>=stealth',shorten >=1pt,thick,roundnode/.style={draw,circle,thick,minimum size=10mm,thick},]
%Nodes
\node[roundnode] (1) {$v_0$};
\node[roundnode] (2) at (2,1.25) [] {$v_1$};
\node[roundnode] (3) [right=of 2] {$v_2$};
\node[roundnode] (4) at (6,1.25) [] {$v_\ell$};
\node[roundnode] (5) at (2,-1.25) [] {$v_{\ell+1}$};
\node[roundnode] (6) [right=of 5] {$v_{\ell+2}$};
\node[roundnode] (7) [right=of 6] {$v_{2\ell}$};
%Lines
\Loop[dist=1cm,dir=NO,label=$0$,labelstyle=above](1)
\Loop[dist=1cm,dir=NO,label=$1$,labelstyle=above](2)
\Loop[dist=1cm,dir=No,label=$2$,labelstyle=above](5)
\draw[->] (1) to [right] node [above] {1} (2);
\draw[->] (2) to [right] node [above] {0} (3);
\node at ($(3)!.5!(4)$) {\ldots};
\draw[->] (4) to [out=100,in=110,looseness=0.9] node [above] {0} (1);
\draw[->] (1) to [right] node [below] {2} (5);
\draw[->] (5) to [right] node [above] {0} (6);
\node at ($(6)!.5!(7)$) {\ldots};
\draw[->] (4) to [out=200,in=80,looseness=0.5] node [above] {2} (5);
\draw[->] (7) to [out=150,in=290,looseness=0.5] node [below] {1} (2);
\draw[->] (7) to [out=240,in=250,looseness=0.6] node [below] {0} (1);
\end{tikzpicture}  
\caption{Graph presentation of the constraint in~\Cref{multiplePeriodic}.}
\label{multiplePeriodicGraph}
\end{figure}
From the structure of the graph, for any sequence $\bfy'$ of length $n-(\ell-1)$, when $n\geq \ell-1$, that is generated by the graph, it holds that ${\bfy\triangleq\bfy'\circ0^{\ell-1}\in F_n(\underline{\bfalpha})\subseteq S(n)}$. So, $|S(n-(\ell-1))|\leq|F_n(\underline{\bfalpha})|\leq|S(n)|$. As a result, $\mathsf{cap}(\underline{\bfalpha})= \mathsf{cap}(S)$.
The adjacency matrix $A$ of this graph satisfies that $A_{i,j}=1$ for $(i,j)$ such that: $i=0, j=0,1,\ell+1$ or $i=j=1$ or $i=j=\ell+1$ or $i=j-1$ if $i\neq2\ell$ or $i=\ell,2\ell ,j=0$ or $i=2\ell, j=1$. Otherwise, $A_{i,j}=0$. 
It can be shown that the characteristic polynomial of this matrix is $x^{\ell-1}(x-1)(x^{\ell+1}-2x^\ell+x^{\ell-1}-2)-(x+1)$. Thus, the capacity of $S$ is $\log_2(\lambda)$ when $\lambda$ is
the largest real root of this polynomial.
\end{IEEEproof}

\section{The Minimal Number of Labels Problem}
\label{sec:minimal}
In this section, we solve Problem~\ref{prob2} for $\ell=1,2$ and any $q$. That is, we find $s(\ell,q)$ which is the minimal number of labels of length $\ell$ that are needed in order to gain labeling capacity of $\log_2(q)$.

\begin{theorem}
    It holds that $s(1,q)=q-1$.
\end{theorem}

\begin{IEEEproof}
Let $\underline{\bfalpha}$ be a vector of $q-1$ labels of length one. The valid $\underline{\bfalpha}$-labeling sequences are all the sequences over $\Sigma_q^n$. So, $\mathsf{cap}(\underline{\bfalpha})=\log_2(q)$. Additionally, let $\underline{\bfbeta}$ be a vector of $q-2$ labels of length one. The valid $\underline{\bfbeta}$-labeling sequences are all the sequences over $\Sigma_{q-1}^n$. So, $\mathsf{cap}(\underline{\bfbeta})=\log_2(q-1)<\log_2(q)$.    
\end{IEEEproof}

Our main result in this section is finding the value of $s(2,q)$. It is easy to see that $s(2,1)=0$. In order to show the result for $q>1$, first, consider the following definition and theorems that will be used in deriving $s(2,q)$.
\begin{definition}
    Let $G=(V,E)$ be a directed graph. This graph is said to be a $\emph{\textbf{path unique graph}}$ if for any $k\geq1$, between any two vertices there exists at most one path of length $k$.
\end{definition}
\begin{example}
The graph in~\Cref{pathUnique} is a path unique graph since there are no paths from $v$ to $u$ or from $v$ to itself, and every different path from $u$ to $v$ or from $u$ to itself is of different length. 
The graph in~\Cref{notPathUnique} is not a path unique graph since there are two paths of length two from $u$ to $v$.

\begin{figure}[h]
\begin{subfigure}{0.5\textwidth}
  \centering
\begin{tikzpicture}[->,>=stealth',shorten >=1pt,thick,round/.style={draw,circle,thick,minimum size=0.5cm,thick},]
    \node[round] at (0,0) (1) [] {$u$};
    \node[round] (2) at (2,0) [] {$v$};
    \Loop[dist=1cm,dir=NO](1)
    \draw[->] (1) to [right] node [above] {} (2);
\end{tikzpicture}
  \caption{}
  \label{pathUnique}
\end{subfigure}
\begin{subfigure}{.1\textwidth}
  \centering
  \begin{tikzpicture}[->,>=stealth',shorten >=1pt,thick,round/.style={draw,circle,thick,minimum size=0.5cm,thick},]
    
    \node[round] at (0,0) (1) [] {$u$};
    \node[round] (2) at (2,0.5) [] {};
    \node[round] (3) at (4,0) [] {$v$};
    \node[round] (4) at (2,-0.5) [] {};
    
    \draw[->] (1) to [right] node [above] {} (2);
    \draw[->] (2) to [right] node [above] {} (3);
    \draw[->] (1) to [right] node [above] {} (4);
    \draw[->] (4) to [right] node [above] {} (3);
    \end{tikzpicture}
  \caption{}
  \label{notPathUnique}
\end{subfigure}
\caption{Examples for a path unique graph (a) and a graph which is not path unique (b).}
\label{fig:fig}
\end{figure}
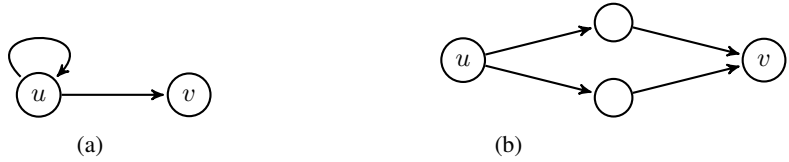
\end{example}
The study of path unique graphs was conducted in~\cite{HUANG20122203}, where they presented the following result.
\begin{theorem}[\cite{HUANG20122203} ] \label{maxSizePathUniqe}
    Let $n$ be a positive integer and let $h(n)$ denote the maximum number of edges in a path unique graph with $n$ vertices. It holds that, 
    $$h(n)= \begin{cases}
    \frac{(n+1)^2}{4} & n \text{ is odd,}\\
    \frac{n(n+2)}{4} & n \text{ is even},
    \end{cases}$$
    and there exists a path unique graph with $h(n)$ edges for any $n$. 
\end{theorem}
The following theorem will show the connection between path unique graphs with $q$ vertices and the value of $s(2,q)$.

\begin{theorem}\label{graphEquivalence}
    Let $\cS$ be a subset of the labels of length two. Let $G=(V,E)$ be a directed graph in which $V=\Sigma_q$ when $q>1$ and $E= \{(x,y)| xy\in \cS\}$. Denote $\bar\cS= \{\bfalpha_1,\ldots,\bfalpha_{q^2-|\cS|}\}$ the set of labels of length two which are not in $\cS$ and $\underline\bfalpha=(\bfalpha_1,\ldots,\bfalpha_{q^2-|\cS|})$. It holds that $\mathsf{cap}(\underline\bfalpha)=\log_2(q)$ if and only if $G$ is a path unique graph.
\end{theorem}

\begin{IEEEproof}
    Let $\cS$ be a subset of labels of length two, let $\bar\cS= \{\bfalpha_1,\ldots,\bfalpha_{q^2-|\cS|}\}$ be the set of labels of length two which are not in $\cS$ and $\underline\bfalpha=(\bfalpha_1,\ldots,\bfalpha_{q^2-|\cS|})$. Let $G=(V,E)$ be a directed graph in which $V=\Sigma_q$ and $E= \{(x,y)| xy\in \cS\}$. Additionally, for $\boldsymbol{x}\in\Sigma_q^n$, let $\boldsymbol{y}\in\Sigma_{1+|\bar{\cS}|}^n$ be the $\underline\bfalpha$-labeling sequence of $\boldsymbol{x}$, so, for $ 1\leq j\le q,\ y_i = j$ if and only if ${\boldsymbol{x}}_{[i;2]} = \bfalpha_j$. 
    
    First, assume that $G$ is a path unique graph. Note that from the definition of path unique graphs, $G$ is not a full graph, which means that $\bar\cS$ is not an empty set. Let $ab\in\bar\cS$. In order to prove that $\mathsf{cap}(\underline\bfalpha)=\log_2(q)$, it will be shown that the capacity of the sequences that start and end with $ab$ is $\log_2(q)$. Moreover, it will be shown that using sequences of length~$n$ that start and end with $ab$, the mapping $L_{\underline\bfalpha}$ is one-to-one, which means that the number of valid $\underline{\bfalpha}$-labeling sequences of length $n$ is at least $q^{n-4}$ and the labeling capacity is $\mathsf{cap}(\underline\bfalpha) =  \limsup\limits_{n\to\infty}\frac{\log_2(q^{n-4})}{n}=\log_2(q)$.

    Let $\bfx_1, \bfx_2\in\Sigma_q^n$ be sequences that start and end with $ab$ such that $\bfy=L_{\underline\bfalpha}(\bfx_1)=L_{\underline\bfalpha}(\bfx_2) \neq (0,\ldots,0)$. It will be shown that $\bfx_1=\bfx_2$.
    If $y_i=j$ for $j>0$, from the definition of $\underline\bfalpha$-labeling sequences, it holds that $\bfx_{1[i;2]} = \bfx_{2[i;2]}= \bfalpha_j$. Otherwise, assume $y_i=0$ and denote by $i_\ell$ the largest index such that $i_\ell< i$, and $y_{i_\ell}\neq0$. Additionally, let $i_r$ be the smallest index such that $i_r> i$, and  $y_{i_r}\neq0$. Such $i_\ell,i_r$ always exist because $\bfx_1,\bfx_2$ start and end with $ab$ when $ab\in\overline{\cS}$. From the definition of $\underline\bfalpha$-labeling, it holds that $\bfx_{1[i';2]}\in \cS$  for $i_\ell\leq i' < i_r$. There is only one path in $G$ of length $m=i_r-i_\ell$ between any two vertices. And so, $\bfx_{1[i_\ell;m]}$ is uniquely determined, so $\bfx_{1[i_\ell;m]}= \bfx_{2[i_\ell;m]}$.

    On the other direction, assume that $G$ is not path unique. So, there exist two vertices $u,v\in V$ with two different paths between $u$ and $v$ of the same length $m>1$. Denote these two paths by $w_1=ut_1t_2\cdots t_{m-1} v$ and $w_2=us_1s_2\cdots 
    s_{m-1}v$ when $t_i,s_i\in V$ for $1\leq i \leq m-1$, $(t_{i},t_{i+1}),(s_{i},s_{i+1})\in E$ for $1\leq i \leq m-2$, and $(u,t_1),(t_{m-1},v),(u,s_1),(s_{m-1},v)\in E$. From the definition of $E$, $t_{i}t_{i+1},s_{i}s_{i+1}\in \cS$ for $1\leq i \leq m-2$, and $ut_1,t_{m-1}v,us_1,s_{m-1}v\in \cS$. So, $L_{\underline\bfalpha}(w_1)= L_{\underline\bfalpha}(w_2)=(0,\ldots,0)$. In other words, the function $L_{\underline\bfalpha}(\cdot)$ does not distinguish between the substrings $\bfw_1$ and $\bfw_2$, when $\bfw_1$ and $\bfw_2$ correspond to the paths $w_1$, $w_2$ respectively.

    Denote the set of all sequences of length $n$ over $\Sigma_q$ which do  not contain $\bfw_2$ as a substring by $$\cK_{\bfw_2}=\{\bfw\in\Sigma_q^n| \bfw \neq \bfp \bfw_2\bfq, \bfp,\bfq\in\Sigma_q^*\}$$. Let $L^*_{\underline\bfalpha}:\cK_{\bfw_2}\to\Sigma^n_{|\cS|+1}$ be a function for which $\forall \bfw\in\cK_{\bfw_2}, L^*_{\underline\bfalpha}(\bfw)=L_{\underline\bfalpha}(\bfw)$. Let $\bfw'\in\Sigma_q^n\setminus \cK_{\bfw_2}$ be a sequence that contains $\bfw_2$ as a substring. Let $\bfw^*\in\Sigma_q^n$ be such that $\bfw^*_{[i;m+1]}=\bfw_1$ if  $\bfw'_{[i;m+1]}=\bfw_2$ and $w^*_i=w'_i$ otherwise. It holds that $\bfw^*\in\cK_{w_2}$ and so $L_{\underline\bfalpha}(\bfw')=L_{\underline\bfalpha}(\bfw^*)=L^*_{\underline\bfalpha}(\bfw^*)$. As a result, $|\text{Im}(L_{\underline\bfalpha})|=|\text{Im}(L^*_{\underline\bfalpha})|\leq |\cK_{\bfw_2}|$, where $\text{Im}(f)$ is the image of $f$. Since the length of $\bfw_2$ is fixed, it holds that $\mathsf{cap}(\cK_{\bfw_2})<\log_2(q)$, and thus $\mathsf{cap}(\underline\bfalpha)< \log_2(q)$. 
\end{IEEEproof}

We are now ready to find the value of $s(2,q)$, which is established using the last two theorems.

\begin{corollary}\label{s2}
    It holds that $s(2,q)=q^2-h(q)$ when $$h(q)= \begin{cases}
    \frac{(q+1)^2}{4} & q \text{ is odd}\\
    \frac{q(q+2)}{4} & q \text{ is even}.
\end{cases}$$
\end{corollary}

\begin{IEEEproof}
For $q=1$, $s(2,1)=1-1=0$. For $q>1$, from~\Cref{maxSizePathUniqe}, the maximal number of edges in a path unique graph with $q$ vertices is $h(q)$, when $h(q)$ is as defined in the corollary. Moreover, there exists a path unique graph with this number of edges for any $q>1$. So, from~\Cref{graphEquivalence}, there are at least $q^2-h(q)$ labels of length two that are needed in order to gain labeling capacity of $\log_2(q)$ and there is a selection of exactly $q^2-h(q)$ labels such that the capacity is $\log_2(q)$.  
\end{IEEEproof}

The next example will show a set of ten labels of length two over $\Sigma_4$ that gain labeling capacity two.

\begin{example}
   Consider the set of the following six labels: $\cS= \{AA, AC, AT, GG, GC, GT\}$ and $\underline\bfalpha=(\bfalpha_1,\ldots,\bfalpha_{10})$ when $\forall i\in[10], \bfalpha_i\in\bar \cS$. Let $G= (V,E)$ be a directed graph in which $V=\Sigma_4$ and $E= \{(x,y)| xy\in \cS\}$. Since $G$ is a path unique graph, we conclude from~\Cref{graphEquivalence} that $\mathsf{cap}(\underline\bfalpha)=2$. 

\begin{figure}[h]
\centering
    \begin{tikzpicture}[->,>=stealth',shorten >=1pt,thick]
    \SetGraphUnit{2} 
    \tikzset{VertexStyle/.style = {draw,circle,thick,minimum size=8mm},thick} 
    \Vertices{line}{A,C,G,T}
    \Loop[dist=1cm,dir=NO](A)
    \Loop[dist=1cm,dir=NO](G)
    \draw[->] (A) to [right] node [above] {} (C);
    \draw[->] (A) to [bend right] node [above] {} (T);
    \draw[->] (G) to [right] node [above] {} (C);
    \draw[->] (G) to [right] node [above] {} (T);
\end{tikzpicture}
\caption{A path unique graph that represents a set of labels that gain labeling capacity of two.}
\end{figure}
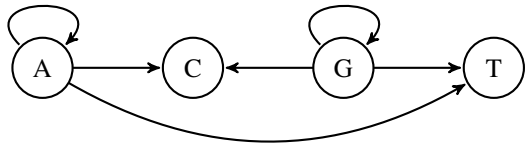  

\end{example}

\section{The Largest Achievable Labeling Capacity for any Number of Labels}\label{sec:largestCap}
This section provides several results and bounds for the case of using any number of labels of length two. From~\Cref{s2}, when using ten or more labels of length two, it is possible to achieve the maximal labeling capacity for the quaternary alphabet. From~\Cref{Largest capacity}, for any alphabet, the largest labeling capacity using a single label of length two is obtained, for example, by the label $AA$ and is equal to $\log_2(1.7549)\approx0.811$ for $q>1$ and $0$ if $q=1$. In order to generalize these results for arbitrary numbers of labels, the following definition will be used.
\begin{definition}
Let $t(k,\ell,q)$ be the maximal labeling capacity that can be achieved using $k$ labels of length $\ell$ above $\Sigma_q$, that is,  $t(k,\ell,q)\triangleq \max{\{\mathsf{cap}(\underline{\bfalpha})| \underline{\bfalpha}\in\Sigma_{q^\ell}^k\}}.$ 
\end{definition}
The case of $k=1$ was studied in~\Cref{Largest capacity}. From~\Cref{s2}, for $q=2$, $t(2,2,q)= \log_2(q)$. In the next theorem we solve the case of two labels for $q\geq 3$.

\begin{theorem}
    Let $q\geq 3$ and $a \neq b\in \Sigma_q$. It holds that $t(2,2,q) = \log_2(\lambda)$ where $\lambda \approx 2.206$ is the largest real root of $x^5-3x^4+3x^3-3x^2+x-1$. Moreover, the pairs of labels that achieve this labeling capacity are only from one of the following types:
    \begin{itemize}
        \item $aa,ab$, i.e., one periodic with period one and one which is non-cyclic, with overlap between them.
        \item $aa,bb$, i.e., both periodic with period one.
        \item $ab,ba$, i.e., both non-cyclic with overlap from both sides.
    \end{itemize}
\end{theorem}

\begin{IEEEproof}
    Let $a,b,c\in \Sigma_q$. The pair of labels could be from the following types:
    \begin{enumerate}
        \item Both non-cyclic with no overlap. For example $\underline{\bfalpha}^1=(ab,ac)$.
        \item Both non-cyclic with overlap from one side. For example $\underline{\bfalpha}^2=(ab,bc)$.
        \item Both non-cyclic with overlap from both sides. For example $\underline{\bfalpha}^3=(ab,ba)$.
        \item Both periodic with period one. For example $\underline{\bfalpha}^4=(aa,bb)$.
        \item One periodic with period one and one which is non-cyclic, with no overlap between them. For example $\underline{\bfalpha}^5=(aa,bc)$.
        \item One periodic with period one and one which is non-cyclic, with overlap between them. For example $\underline{\bfalpha}^6=(aa,ab)$.
    \end{enumerate}
    Using each pair of labels will result in a labeling sequence, which is a ternary sequence that satisfies some constraints. In order to compare between the different options for pairs of labels,~\Cref{table:2} summarizes the subsequences that cannot appear in the labeling sequences in each case.
\begin{table}[h!]
\centering
\caption{The subsequences that can not appear in the labeling ternary sequences using different types of two labels.}
\begin{tabular}{|p{4cm}|p{1cm}|p{2cm}|} 
 \hline
 Labels properties & Example & forbidden subsequences \\ [0.5ex] 
 \hline\hline
 Both non-cyclic, no overlap & $ab,ac$ & $11,12,21,22$ \\ 
 Both non-cyclic, one-side overlap & $ab,bc$ & $11,21,22$ \\
 Both non-cyclic, two-sides overlap & $ab,ba$ & $11,101,22,202$ \\
 Both periodic, period 1 & $aa,bb$ & $12,101,21,202$ \\ 
 Periodic and non-cyclic, no overlap & $aa,bc$ & $12,101,21,22$ \\
 Periodic and non-cyclic, one-side overlap & $aa,ab$ & $101,102,21,22$ \\
 [1ex] 
 \hline
\end{tabular}
\label{table:2}
\end{table}

Next, note that the sequences that can be generated by the graph in~\Cref{ternaryGraph} are all the ternary sequences. As expected, the capacity of the constraint that is presented in this graph is $\log_2(3)$.
\begin{figure}[h]
\centering
    \begin{tikzpicture}[->,>=stealth',shorten >=1pt,thick,roundnode/.style={draw,circle,thick,minimum size=10mm,thick},]
%Nodes
\node[roundnode] (1) {$v_0$};
\node[roundnode] (2) at (2,1.25) [] {$v_1$};
\node[roundnode] (3) [right=of 2] {$v_1'$};
\node[roundnode] (4) at (2,-1.25) [] {$v_2$};
\node[roundnode] (5) [right=of 5] {$v_2'$};
%Lines
\Loop[dist=1cm,dir=NO,label=$0$,labelstyle=above](1)
\Loop[dist=1cm,dir=NO,label=$1$,labelstyle=above, color=red](2)
\Loop[dist=1cm,dir=SO,label=$2$,labelstyle=below, color=red](4)
\draw[->] (1) to [right] node [above] {1} (2);
\draw[->] (2) to [right] node [below] {0} (3);
\draw[->] (3) to [out=100,in=110,looseness=1.1] node [above] {0} (1);
\draw[->] (1) to [right] node [below] {2} (4);
\draw[->] (4) to [right] node [above] {0} (5);
\draw[->, color=red] (3) to [out=270,in=80,looseness=0.5] node [above] {2} (4);
\draw[->, color=red] (5) to [out=150,in=290,looseness=0.5] node [below] {1} (2);
\draw[->] (5) to [out=240,in=250,looseness=1.3] node [below] {0} (1);
\draw[->, color=red] (5) to [out=210,in=330,looseness=0.6] node [below] {2} (4);
\draw[->, color=red] (3) to [out=140,in=30,looseness=0.65] node [above] {1} (2);
\draw[->, color= red] (2) to [right] node [right] {2} (4);
\draw[->, color=red] (4) to [out=110,in=250,looseness=0.6] node [left] {1} (2);
\end{tikzpicture}  
\caption{Graph presentation of all the ternary sequences.}
\label{ternaryGraph}
\end{figure}
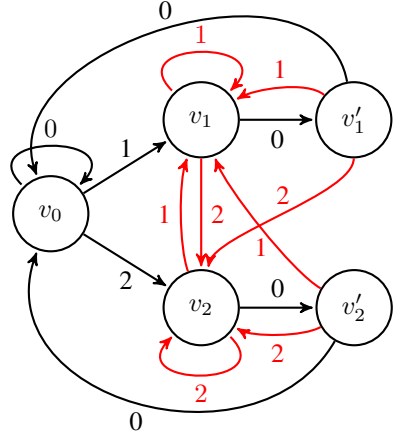

The adjacency matrix of this graph is
$$\begin{pmatrix}
1 & 1 & 0 & 1 & 0\\
0 & \textcolor{red}{1} & 1 & \textcolor{red}{1} & 0\\
1 & \textcolor{red}{1} & 0 & \textcolor{red}{1} & 0\\
0 & \textcolor{red}{1} & 0 & \textcolor{red}{1} & 1\\
1 & \textcolor{red}{1} & 0 & \textcolor{red}{1} & 0\\
\end{pmatrix}.$$

Adding constraints to the set of all ternary sequences will cause the removal of some of the red edges. Looking at the above adjacency matrix, each 1 in red represents a subsequence which is forbidden to use in some of the six cases of using a pair of labels. For example, the $1$ in entry $(1,1)$ (the second row and second column) will be turned to a $0$ if the subsequence $11$ will be forbidden which happens for $\underline{\bfalpha}^1,\underline{\bfalpha}^2,$ and $\underline{\bfalpha}^3$. The adjacency matrix of the graphs that present the constraints using pair of labels can be presented as the following matrix, in which each non-binary cell (highlighted in red) is $0$ if the subsequence in that cell is forbidden:
$$\begin{pmatrix}
1 & 1 & 0 & 1 & 0\\
0 & \textcolor{red}{11} & 1 & \textcolor{red}{12} & 0\\
1 & \textcolor{red}{101} & 0 & \textcolor{red}{102} & 0\\
0 & \textcolor{red}{21} & 0 & \textcolor{red}{22} & 1\\
1 & \textcolor{red}{201} & 0 & \textcolor{red}{202} & 0\\
\end{pmatrix}.$$ 
For example, the adjacency matrix of the graph presentation of the constraint that is obtained using $\underline{\bfalpha}^1$ is:
$$\begin{pmatrix}
1 & 1 & 0 & 1 & 0\\
0 & \textcolor{red}{0} & 1 & \textcolor{red}{0} & 0\\
1 & \textcolor{red}{1} & 0 & \textcolor{red}{1} & 0\\
0 & \textcolor{red}{0} & 0 & \textcolor{red}{0} & 1\\
1 & \textcolor{red}{1} & 0 & \textcolor{red}{1} & 0\\
\end{pmatrix}.$$

After going through all the options for pairs of labels, the largest capacity is obtained for the graphs that present constraints that has the same capacity as the labeling capacity of the third, forth and sixth cases. 

For the completeness of the proof, the labeling capacity for each case is presented next. From~\Cref{multiplePeriodic}, the labeling capacity for the fourth case (and so, for the third and sixth cases) is $\log_2(\lambda)$ where $\lambda \approx 2.206$ is the largest real root of $x^5-3x^4+3x^3-3x^2+x-1$. From~\Cref{multipleNonOverlaping}, $\mathsf{cap}(\underline{\bfalpha}^1)=\log_2(\lambda_1)$ where $\lambda_1=2$ is the largest real root of $x^2-x-2$. From~\Cref{twoLabelsTheorem}, $\mathsf{cap}(\underline{\bfalpha}^2)=\log_2(\lambda_2)$, where $\lambda_2\approx 2.148$ is the largest real root of $x^3-x^2-2x-1$. Using the matrix and the table, $\mathsf{cap}(\underline{\bfalpha}^5)=\log_2(\lambda_5)$ where $\lambda_5\approx2.107$ is the largest real root of $x^4-2x^3-1$.
\end{IEEEproof}

As is shown in~\Cref{s2}, for the quaternary alphabet, the maximal labeling capacity can be gained using ten or more labels of length two. Lastly, a bound for the case of using nine labels of length two for the any alphabet is given. The proof is based on the idea of using a graph which is almost path unique.
\begin{theorem}
    For $q\geq 4$, it holds that $t(9,2,q) \geq \log_2(3.866)$.
\end{theorem}
\begin{IEEEproof}
    First, from~\Cref{s2}, for $q<4$ it holds that $t(9,2,q) = \log_2(q)$. For $q=4$, for convenience, assume $\Sigma_4=\{A,C,G,T\}$. Let $\cS=\{AC,CA,GA,GC,GG,TA,TC,TG,TT\}$ be the set of nine labels of length two that will be in use. As was demonstrated in~\Cref{graphEquivalence}, the labels that are not in use can be presented in a graph as is shown in~\Cref{nineLabelsGraph}.
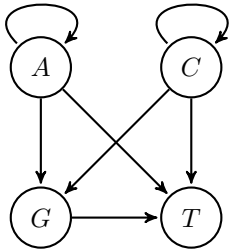
\begin{figure}[h]
\centering
    \begin{tikzpicture}[->,>=stealth',shorten >=1pt,thick]
    \SetGraphUnit{2} 
    \tikzset{round/.style= {draw,circle,thick,minimum size=8mm},thick} 
    \node[round] at (0,0) (A) [] {$A$};
    \node[round] (C) at (2,0) [] {$C$};
    \node[round] (G) at (0,-2) [] {$G$};
    \node[round] (T) at (2,-2) [] {$T$};
    \Loop[dist=1cm,dir=NO](A)
    \Loop[dist=1cm,dir=NO](C)
    \draw[->] (A) to [right] node [above] {} (G);
    \draw[->] (A) to [right] node [above] {} (T);
    \draw[->] (C) to [right] node [above] {} (G);
    \draw[->] (C) to [right] node [above] {} (T);
    \draw[->] (G) to [right] node [above] {} (T);
\end{tikzpicture}
\caption{A graph that represents a set of seven labels that are not being used.}
\label{nineLabelsGraph}
\end{figure}  
    This graph was generated from a path unique graph that the edge $GT$ was added to. Note that this is not a path unique graph, since it has four vertices and more than six edges. There are two  ways to get from $A$ to $T$ in two steps: $A\rightarrow A \rightarrow T$, $A\rightarrow G \rightarrow T$, and there are two ways to get from $C$ to $T$ in two steps: $C \rightarrow C \rightarrow T$, $C \rightarrow G \rightarrow T$. It was shown in~\Cref{graphEquivalence} that if the labels that are not being used generate a path unique graph, the maximal labeling capacity is being obtained. So, in this case, assume that we do not use all of the sequences over $\Sigma_4$, but only the sequences that do not contain $AGT,CGT$ as substrings. Note that in this way, it is not possible to get from $A$ or $C$ to $T$ in two different ways with the same number of steps (and not only for $2$ steps). The constraint of not having the sequences $AGT,CGT$ as substrings in a sequence over $\Sigma_4$ is presented in~\Cref{withoutAGTcGT}.
\begin{figure}[h]
\centering
    \begin{tikzpicture}[->,>=stealth',shorten >=1pt,thick,round/.style={draw,circle,thick,minimum size=1cm,thick},]
\SetGraphUnit{3} 
\node[round] at (1,0) (A) [] {$v_0$};
\node[round] (B) at (4,0) [] {$v_1$};
\node[round] at (7,0) (C) [] {$v_2$};
\node[round] (D) at (4,-2) [] {$v_3$};
\Loop[dist=1cm,dir=NO,label=${G,T}$,labelstyle=above](A)
\Loop[dist=1cm,dir=NO,label=${A}$,labelstyle=above](B)
\Loop[dist=1cm,dir=SO,label=${C}$,labelstyle=below](D)
\draw[->] (A) to [right] node [above] {$A$} (B);
\draw[->] (B) to [bend right] node [above] {$T$} (A);
\draw[->] (B) to [right] node [above] {$G$} (C);
\draw[->] (C) to [bend right] node [above] {$A$} (B);
\draw[->] (A) to [bend right] node [above] {$C$} (D);
\draw[->] (D) to [right] node [above] {$T$} (A);
\draw[->] (C) to [bend left] node [above] {$C$} (D);
\draw[->] (D) to [right] node [above] {$G$} (C);
\draw[->] (B) to [bend right] node [left] {$C$} (D);
\draw[->] (D) to [bend right] node [right] {$A$} (B);
\draw[->] (C) to [out=120,in=55] node [above] {$G$} (A);
\end{tikzpicture} 
\caption{A graph presentation of the constraint of not having the subsequences $AGT,CGT$.}
\label{withoutAGTcGT}
\end{figure}
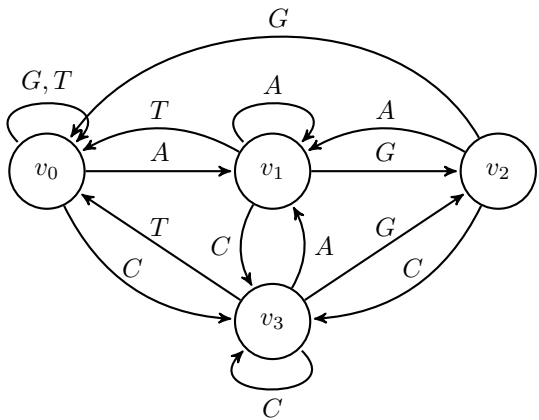
The adjacency matrix of this graph is
$$\begin{pmatrix}
2 & 1 & 0 & 1\\
1 & 1 & 1 & 1\\
1 & 1 & 0 & 1\\
1 & 1 & 1 & 1\\
\end{pmatrix},$$ and its characteristic polynomial is $x^4-4x^3+2x$. Thus, the capacity of this constraint is $\log_2(\lambda)$ when $\lambda\approx3.866$ is the largest real root of this polynomial. Lastly, for $q>4$, assume $\Sigma_4\subseteq \Sigma_q$ and let $\cS$ be the set of nine labels that is used for $q=4$. As was shown for the case of $q=4$, assume that we do not use all of the sequences over $\Sigma_q$, but only the sequences which are above $\Sigma_4$ that do not contain $AGT, CGT$ as substrings.
\end{IEEEproof}

\section{Conclusion}
In this work, we modeled the process of DNA labeling and provided the labeling capacity that is gained using one or more labels. This analysis helped us organizing the labels of lengths $\ell\leq 5$ by their labeling capacity. Moreover, we showed a connection between path uniqueness of a graph and the ability of achieving the full labeling capacity using labels of length two over any alphabet. It led us to discuss the maximal labeling capacity that can be achieved using an arbitrary number of labels of length two. In this work we showed the result for two labels and a lower bound on the case of using nine labels when $q=4$. For three labels, a lower bound on this value can be obtained using labels of the form $\underline{\bfalpha}=(aa,bb,ab)$ when $a,b\in \Sigma_q, q\geq3$. So, $t(3,2,q) \geq \log_2(\lambda)$ when $\lambda\approx2.582$ is the largest real root of $x^5-3x^4+2x^3-3x^2+2x-1$. The exact value of the maximal labeling capacity for any number of labels of any length over any alphabet, and the labels that can achieve this value are still need to be studied.

\section{Acknowledgement}
The authors thank Ron M. Roth and Christoph Hofmeister for helpful discussions and for pointing us to the results in \cite{HUANG20122203}.
    
\bibliographystyle{IEEEtran}
\bibliography{refs}
%\printbibliography

\end{document}